\documentclass[11pt]{article}

\usepackage[usenames,dvipsnames]{xcolor}
 \usepackage{import}
 \usepackage{float}
 \usepackage{calc}
 \usepackage{booktabs}
 \usepackage{amsfonts}
 \usepackage{latexsym}
 \usepackage{placeins}
 \ifx\pdftexversion\undefined
   \usepackage[dvips]{graphicx}
 \else
   \usepackage[pdftex]{graphicx}
 \fi

 \usepackage{subcaption}
 \usepackage{multirow}
 \usepackage{multicol}
 \usepackage{amssymb}
 \usepackage{xfrac}
 \usepackage{arydshln}
 \usepackage{hyperref}  
 \usepackage[top=1.2in, bottom=2in, left=1in, right=1in]{geometry}

 \usepackage{array}

 \usepackage{adjustbox}
 \usepackage{fancyref}

 \usepackage{enumitem}

 \usepackage{authblk}
 \usepackage{amsmath}
 \usepackage[cp1250]{inputenc}
 \usepackage[OT4]{fontenc}
 \usepackage[round]{natbib}
 \let\LaTeXtitle\title
 \renewcommand{\title}[1]{\LaTeXtitle{\huge#1}}

 \usepackage{csquotes}

 \usepackage{xpatch}
 \xpatchcmd{\author}{\relax#1\relax}{\relax\detokenize{#1}\relax}{}{}

\begin{document}

\title{L-moments for automatic threshold selection in extreme value analysis}
\date{}

\author[\empty]{Jessica Silva Lomba\textsuperscript{1,}\thanks{Corresponding author. E-mail: \texttt{jslomba@fc.ul.pt}}}
\author[1]{Maria Isabel Fraga Alves}
\affil[1]{CEAUL and DEIO, Faculty of Sciences, University of Lisbon}


\maketitle
\vspace{-0.8cm}
\begin{abstract}

In extreme value analysis, sensitivity of inference to the definition of extreme event is a paramount issue. Under the peaks-over-threshold (POT) approach, this translates directly into the need of fitting a Generalized Pareto distribution to observations above a suitable level that balances bias versus variance of estimates. Selection methodologies established in the literature face recurrent challenges such as an inherent subjectivity or high computational intensity. We suggest a truly automated method for threshold detection, aiming at time efficiency and elimination of subjective judgment. Based on the well-established theory of L-moments, this versatile data-driven technique can handle batch processing of large collections of extremes data, while also presenting good performance on small samples.

\noindent The technique's performance is evaluated in a large simulation study and illustrated with significant wave height data sets from the literature. We find that it compares favorably to other state-of-the-art methods regarding the choice of threshold, associated parameter estimation and the ultimate goal of computationally efficient return level estimation.
\end{abstract}


\section{Introduction}

Extreme Value Theory (EVT) provides a statistical framework focused on explaining the behaviour of observations that fall furthest from the central bulk of data from a given phenomenon. Seismic magnitudes, sea or precipitation levels, the prices of financial assets are examples of traditional subjects of extreme value analysis (EVA), for which the greatest interest and the greatest risk lies in the most rare and extreme events. Given the nature of these processes, accurate inference can aid in preventing serious consequences, but this requires appropriate definition of what are extreme observations.

A common approach is to perform estimation using the observations above a previously chosen level -- from a parametric standpoint, this corresponds to the peaks-over-threshold (POT) approach. The methodology is based on the asymptotic approximation to the Generalized Pareto distribution (GPd) of $Y:=X-u$, the rescaled excesses above a suitably high level $u$, should the non-degenerate limiting distribution exist -- \cite{pickands1975}. 

The cumulative distribution function of the GPd, with shape parameter $\xi\in\mathbb{R}$ -- known as the Extreme Value Index (EVI) --, threshold-dependent scale parameter $\sigma_u>0$ and location at the origin takes the form

\begin{equation}\label{GP}
  H_{\xi}(y|\sigma_u):=\left\{
  \begin{aligned}
   &1-\left(1+ \frac{\xi y}{\sigma_u}\right)^{-\sfrac{1}{\xi}}, &y\in (0,\infty),\;\xi>0 \\
   &1-\exp\left(-\frac{y}{\sigma_u}\right), &y\in (0,\infty),\;\xi=0 \\
   &1-\left(1+ \frac{\xi y}{\sigma_u}\right)^{-\sfrac{1}{\xi}}, &y\in (0,-\frac{\sigma_u}{\xi}),\;\xi<0 \\
 \end{aligned}
 \right..
\end{equation}

\vspace{0.2cm}
In practice, the first step in the POT analysis is to choose an appropriate threshold, prior to any type of inference, so we can use the set of observed values from a random sample $X_1,\ldots,X_n$ to estimate $(\xi,\sigma_u)$ under the GPd fit to the excesses above $u$. It is then clear that every inference performed after this point is highly sensitive to the selected cut-off: values of threshold that are too low introduce bias in estimation, since the approximation to the GPd may not hold; on the other hand, high variance of estimators is to be expected when the selected level is too high, since this significantly reduces the dimension of the excesses sample. Finding the balance between these two issues is a task still discussed today, and many suggestions from different perspectives have arisen in the literature -- \cite{scarrottmacdonald2012} and \cite{langousis2016} review some of the methods. We will now briefly mention an assortment of contributions from both frames of classical and Bayesian statistics.

The fastest and most direct ways to choose the threshold are the so-called \emph{rules-of-thumb}, where a fixed number of observations are used (e.g. \citealp{ferreiraetal2003}, suggest $[\sqrt {n}]$ ), but being highly dependent on sample size, this offers no guarantees of suitability for each particular data set. Other methodologies are based on asymptotic properties of estimators for the parameters, such as the asymptotic mean squared error of the Hill estimator for the tail index $\xi$ -- see \cite{danielssonetal2001}, \cite{beirlantetal2002}, \cite{schneideretal2019} or \cite{um2010} for a comparative study under semi-parametric approach; however, asymptotic arguments may not be satisfactorily met when working with small samples, and the combination with bootstrap approaches generally comes with very intense computational effort. Several authors have alternatively considered incorporating the information from the bulk of more central data for determination of the threshold, by presenting mixture models that, for example, combine the flexibility of the GPd tail fit with another appropriate distribution for the data below the threshold; \cite{scarrottmacdonald2012} analyze some examples. A different group of methods is based on evaluating how close is the empirical distribution function to the fitted tail model via distance measures such as the Kolmogorov-Smirnov statistic -- see \cite{clausetetal2009} and \cite{danielssonetal2016} for two different contributions.

Visual diagnostics tools, such as the QQ-plot or the GPd parameter estimates stability plots, are a traditional way of threshold selection -- we can find a brief description in \cite{coles2001}. Recently, new diagnostics plots have been introduced: following the work of \cite{wadsworthtawn2012}, \cite{northropcoleman2014} derive a new multiple-threshold GP model with piece-wise constant shape parameter, which allows for the construction of a plot of p-values from the score test of the hypotheses of shape constancy above each candidate threshold -- the authors suggest that the threshold chosen should correspond to a marked increase in the associated p-values; \cite{wadsworth2016} explored the independent-increments structure of maximum likelihood (ML) estimators, under the non-homogeneous Poisson Process representation, to produce diagnostics plots with more direct interpretability than traditional parameter stability plots; contributions from the field of Bayesian statistics include the works of \cite{leeetal2015} (with recent application by \citealp{manurungetal2018}), who plot measures of surprise at different candidates to select thresholds for uni- and multivariate data sets, and \cite{northropetal2017} who use cross-validation and model averaging to both select a level and provide robustness to the inference against single threshold selection.

Several authors have looked into developing automatic selection methods, in an effort to grant objectivity to such inherently subjective visual tools, and ultimately allow for the simultaneous treatment of large batches of data sets. \cite{thompsonetal2009} developed an automated method based on the approximated normal distribution of the parameter estimates' differences at various thresholds. \cite{langousis2016} use the method of weighted least squares (WLS) to fit linear models to the points above each observation in the traditional mean residual life plot (MRLP) -- the threshold should then be selected as the lowest level corresponding to a local minimum of the weighted mean squared error (WMSE) of the linear fit. \cite{baderetal2018} have built an automatic selection method combining the works of \cite{choulakian2001} on goodness-of-fit (GoF) testing under the GPd hypothesis with the stopping rules for control of the false discovery rate of ordered hypothesis of \cite{gselletal2016}.

\vspace{0.3cm}
In this paper we present a new automatic and computationally effective method for single threshold choice under the POT-GPd methodology. The approach is motivated by, but not limited to, a visual diagnostics tool from the well developed theory of L-moments, using an alternative skewness/kurtosis GPd characterization. Our aim is to fully eliminate the subjectivity of the practitioner from the analysis, presenting a procedure that is fast and simple to apply, allowing for instant and simultaneous threshold selection for large collections of data sets, while maintaining good properties when dealing with small samples.

We give a brief introduction to L-moment theory in Section 2, before exposing and justifying our proposed selection method. Section 3 describes the large scale comparative simulation study performed to validate the methodology against two previously mentioned state-of-the-art automatized procedures. In Section 4 we analyze two significant wave height data sets from the literature, comparing the effect of the chosen threshold across several methods in parameter and return level estimation. Section 5 concludes this work with some final considerations, pointing towards some topics of interest for further investigation.

\section{Automated Threshold Choice}

We now lay the ground for our suggested threshold selection procedure. The aim is to choose the lowest point, from a set of candidates, after which the GP approximation to the sample of excesses is judged to hold sufficiently well. We present an heuristic, automatic and objective adaptation of a known visual diagnostics tool, described for example in \cite{ribatet2011}, and commonly used in the field of Hydrology and Regional Frequency Analysis (RFA) to discriminate between candidate distributions for regional data. The well established fundamental theory of L-moments that underlies the methodology was thoroughly explored by \cite{hosking1986} and a guide to its application to RFA was assembled in \cite{hoskingwallis1997}.

\subsection{L-moment Theory}
Probability weighted moments (PWM) were presented by \cite{greenwood1979} as a parallel theory to that of conventional moments, which provided a way to summarize probability distributions, perform estimation of parameters and hypothesis testing. However, these quantities are hard to interpret in terms of distributional features. \cite{hosking1986} found that specific \emph{linear} combinations of PWM, the \emph{L}-moments, could evade this problem and be read as measures of location, scale and shape of probability distributions, maintaining the good properties of the PWM. 

Consider a random variable $X$ with distribution function $F$: the L-moments present an alternative way of performing PWM-based analysis with increased accuracy and interpretability, allowing for an easier description and identification of distributions and estimation of their parameters. For the entire set of L-moments of $X$ to exist and uniquely define its distribution, it is sufficient that $\mathbb{E}|X|<\infty$, thus describing a wider range of probability distributions than the conventional moments -- comments on this property can be found in \cite{hosking2006}. Given the specific PWM $\alpha_r=M_{1,0,r}=\mathbb{E}\left[X\{1-F(X)\}^r\right]$, we can write the first four L-moments as
\begin{alignat}{2}
 &\lambda_1 = \alpha_0 && \qquad \text{L-location} \equiv \text{expected value} 
    \nonumber\\	
 &\lambda_2 = \alpha_0-2\,\alpha_1 && \qquad \text{L-scale}
    \nonumber\\
 &\lambda_3 = \alpha_0-6\,\alpha_1+6\,\alpha_2 && \qquad \text{third L-moment} 
    \nonumber\\
 &\lambda_4 = \alpha_0-12\,\alpha_1+30\,\alpha_2-20\,\alpha_3 && \qquad \text{fourth L-moment}  
    \nonumber
\end{alignat}
and dimensionless versions, the \emph{L-moment ratios}, as $\tau_r = \sfrac{\lambda_r}{\lambda_2},\; r=3,4,\ldots$ . These ratios satisfy $|\tau_r|<1$, and give measures of distributional shape independently of scale. As such, we make use of
\begin{alignat}{2}
 &\tau_3 = \frac{\lambda_3}{\lambda_2} && \qquad \text{L-skewness}
    \nonumber\\	
 &\tau_4 = \frac{\lambda_4}{\lambda_2} && \qquad \text{L-kurtosis.}
    \nonumber
\end{alignat}
An alternative general bound for the L-kurtosis can be given in terms of L-skewness for any distribution with finite expected value: 
\begin{equation}\label{genbounds}
    \frac{1}{4} \left( 5\,\tau_3^2-1\right )\leq\tau_4 < 1.
\end{equation}

These quantities have been explicitly computed for a number of useful distributions in terms of their parameters -- we can find a list of 16 specific cases in \cite{hosking1986}, including our distribution of interest, the GPd. For a random variable with distribution function \eqref{GP}, the set of L-moments is defined for $\xi<1$, with
\begin{alignat}{3}
 &\lambda_1=\dfrac{\sigma_u}{1-\xi}\;, && \qquad \lambda_2=\dfrac{\sigma_u}{(1-\xi)(2-\xi)}\;, &&
    \qquad \tau_3=\dfrac{1+\xi}{3-\xi}
    \label{lambda1e2etau3GP}
\end{alignat}
and the particular relationship between L-skewness and L-kurtosis
\begin{equation}\label{relationtauGP}
    \tau_4=\tau_3\,\dfrac{1+5\tau_3}{5+\tau_3}\;.
\end{equation}
An alternative way for direct estimation of the parameters of the Generalized Pareto distribution is suggested by these expressions -- see \cite{hoskingwallis1987}. It is also possible to consider the GPd with non-null location: if $X^*=\mu+X$ (a shift in location from the origin to $\mu$) the L-moments for $ X^*$ are the same as for $X$ with the only exception of the L-location $\lambda_1^* = \mu + \lambda_1 $.

Estimation of L-moments is simple and straightforward, with estimators given as linear combinations of the ordered elements of a sample $x_{1:n}\leq x_{2:n}\leq \ldots \leq x_{n:n}$. \cite{hoskingwallis1987} mention several theoretical advantages of these estimators  -- the \emph{L-statistics} -- when compared with the conventional sample moments, namely that they are less subject to estimation bias and also less vulnerable to the effects of sampling variability or measurement errors in extreme data values. Moreover, they allow for more secure inferences from small samples when compared to their classical counterparts, and even to the maximum likelihood estimators in some practical cases. Given the \emph{unbiased} estimators of the $\alpha_r$ PWM 
\begin{equation*}\label{a_r}
    a_r=\frac{1}{n}\,\sum_{i=1}^{n}\binom{n-i}{r}\,x_{i:n}\,\binom{n-1}{r}^{-1}\,,\qquad r=0,1,\ldots,n-1,
\end{equation*}
we build the following set of unbiased estimators for the first four L-moments:
\begin{alignat}{2}
 &\ell_1 = a_0 && \qquad \text{sample mean}
    \nonumber \\	
 &\ell_2 = a_0-2\,a_1 && \qquad \text{sample L-scale}
    \nonumber \\
 &\ell_3 = a_0-6\,a_1+6\,a_2 && \qquad \text{third L-statistic} 
    \nonumber\\
 &\ell_4 = a_0-12\,a_1+30\,a_2-20\,a_3 && \qquad \text{fourth L-statistic}  
    \nonumber
\end{alignat}    
-- \cite{hoskingbalakrishnan2015} addressed the uniqueness and computation of these statistics. In addition, asymptotically unbiased estimators for the first two L-moment ratios are given by
\begin{alignat}{2}   
 &\mathnormal{t}_3 = \frac{\ell_3}{\ell_2} && \qquad \text{sample L-skewness}
    \nonumber \\	
 &\mathnormal{t}_4 = \frac{\ell_4}{\ell_2} && \qquad \text{sample L-kurtosis.}
    \nonumber  
\end{alignat}
Unlike classical sample skewness and kurtosis, these ratios are not algebraically bounded in relation to sample size. Values of $(\mathnormal{t}_3,\mathnormal{t}_4)$ computed from data can challenge the general theoretical bounds in \eqref{genbounds}, although this rarely happens in practical situations.

If the underlying distribution has finite variance, it is possible to demonstrate the asymptotic normality of the estimators $a_r$, $\ell_r$ and $\mathnormal{t}_r$, as well as to compute the corresponding asymptotic bias and variance. In fact, it has been empirically shown that in small samples these estimators approximate their asymptotic normality more closely than traditional sample moments, and that this approximation is often good for samples of dimension as small as $n=20$ \citep{hosking1986}.

\subsubsection*{The L-moment Ratio Diagram (LMRD)}

\begin{figure}[ht!]
  \centering
    \includegraphics[width=0.7\textwidth]{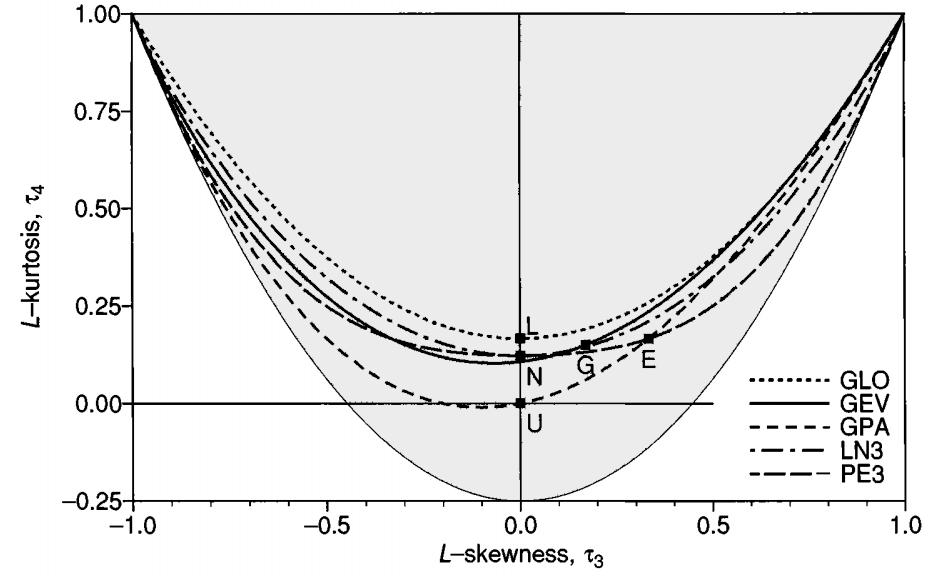}
    \vspace{-0.2cm}
    \caption{\small{L-moment ratio diagram. Figure 2.5, page 25 of \cite{hoskingwallis1997}. The shaded region corresponds to the general bounds on $\tau_3$ and $\tau_4$ according to \eqref{genbounds}.}}
    \label{fig:lmomratiodiagram}
\end{figure}

The L-moment ratio diagram is a visual tool useful for the representation of L-moments from several distributions. In this work, we plot the most commonly used ratios $\tau_3$ and $\tau_4$, a graphical visualization of L-kurtosis as function of L-skewness. Distribution families defined by location and scale parameters appear as a single point -- symmetric distributions have null odd ratios and as such will be plotted on the $\tau_4$ axis. Three-parameter distributions plot as a line: different values of the shape parameter correspond to different points on that line. A particular shape-$\xi$  value  corresponds  to one and only one point $(\tau_3,\tau_4)=\left(\tau_3,g(\tau_3)\right)$, running in the specific curve for the model $H_\xi(y|\mu, \sigma)$, as $\xi$ varies in its range, for whatever location-$\mu$ and scale-$\sigma$ considered. For this representation it is then convenient to find explicit expressions of $\tau_4$ as function of $\tau_3$ such as \eqref{relationtauGP}, specific of the GPd and plotted as the dashed line in Figure \ref{fig:lmomratiodiagram}.

Note that both the Exponential distribution (E -- corresponding to $\xi=0$) and Uniform distribution (U -- corresponding to $\xi=-1$) are plotted on the GPd line, as is to be expected since these are particular cases of the latter. Moreover, having that the Uniform distribution is constant and symmetric, we find it plotted at the origin, meaning it is the only symmetric member of the Generalized Pareto family. From \eqref{lambda1e2etau3GP} we find that negative values of $\tau_3$ are related to underlying shape parameter values $\xi<-1$, corresponding to extremely light tails that seldom appear in practice. As such, it is common to only represent the positive semi-axis of the GPd LMRD.

This graphical tool has been used in RFA, among other purposes, as a visual method for discussing the agreeability of a sample to one or several candidate distributions that could be assumed, by plotting the pair of estimates $(\mathnormal{t}_3,\mathnormal{t}_4)$ and judging its proximity to the proposed theoretical lines. This is the general concept on which we base our automatic threshold selection procedure.

\subsection{Selection Procedure}

The LMRD has been previously used in the context of threshold diagnostics. The \emph{lmomplot} function from the R package \texttt{POT}, described in \cite{ribatet2011}, implements the visual methodology: given a data array \texttt{x}, the estimates $(\mathnormal{t}_{3,u},\mathnormal{t}_{4,u})$ for the sample of excesses over $u$ are computed and plotted on the GPd-specific LMRD, for a set of \texttt{nt} candidate thresholds $u\in$ \texttt{u.range}; by default, the function considers at least 50 equally spaced candidates that cover the range of the original data \texttt{x}. The code also includes a Boolean \texttt{identify} option, which allows for direct identification in the plot of the threshold associated with each estimate point. It is then the user's job to visually decide which of the L-statistics pair is closer to the GPd theoretical line given by \eqref{relationtauGP}, i.e., select the threshold after which the behaviour of these estimators can be considered approximately the expected for the GPd. According to the author, this technique often performed poorly in real data.  

We propose several modifications to this procedure in order to make it automatic and easily applied by practitioners without need for subjective judgment to any data set. Given a sample $x_1,\ldots,x_n$ of size $n$, the Automatic L-moment Ratio Selection Method (ALRSM) works as follows:
\begin{enumerate}
    \item Define the set of candidate thresholds $\left\{u_i\right\}_{i=1}^I$ as one of two reasonable alternatives:
 \begin{eqnarray}
\bullet & I=10  \mbox{ sample quantiles, starting at 25\% by steps of 7.5\%; }  \label{I10} \\
\bullet & I=20  \mbox{ sample quantiles, starting at 25\% by steps of 3.7\%;} \label{I20}
 \end{eqnarray} \vspace{-1cm}
    \item Compute the sample L-skewness and L-kurtosis for the excesses over each candidate threshold $(\mathnormal{t}_{3,u_i},\mathnormal{t}_{4,u_i})$ and determine $d_{u_i}$ -- the minimum Euclidean distance between each point and the GPd theoretical curve \eqref{relationtauGP}:
    \begin{equation*}\label{mindist}
         d_{u_i}=\min_{\tau_3}\sqrt{(t_{3,u_i}- \tau_3)^2+(t_{4,u_i}-g(\tau_3))^2}, \,\,\mbox{for}\,\, i=1,\cdots, I, 
    \end{equation*}
    with
    \begin{equation*} 
        g(\tau_3):=\tau_3\,\frac{1+5\tau_3}{5+\tau_3};
    \end{equation*}
    \item The threshold after which the behaviour of the tail of the underlying distribution can be considered approximately Generalized Pareto is then automatically selected as \begin{equation}\label{u_star}
        u^*= \operatorname*{arg\;min}_{u_i, \,1\leq i \leq I}\left\{d_{u_i} \right\}\,,
    \end{equation} that is, the level above which the corresponding L-statistics fall closest to the curve.
\end{enumerate}

\newpage
The main differences when comparing this method to the previously mentioned \emph{lmomplot} function lie in the definition of the candidates' set and the mathematical (thus automatic) evaluation of the closeness of estimates to theoretical values. The choice of a smaller set with only 10 or 20 candidates is common in the cited references in Section 1; furthermore, by only evaluating thresholds higher than the 25\% sample quantiles we avoid selecting levels that are, in the majority of cases, too low for a good approximation to the GPd; the very high levels above the 92.5\%/95.3\% sample quantiles are also cast away from the evaluation according to choice of candidates in \eqref{I10} and \eqref{I20}, in an attempt to guarantee a sufficient amount of excesses for acceptable estimation of the L-statistics.

Although this procedure is fundamentally heuristic in nature, we found that its performance is often satisfying both for simulated samples and real data sets, in terms of accuracy of the threshold selected and consequent quality of parameter and return level estimation. Given the simplicity of the computational processes involved, the method selects a level almost instantly, making it very convenient for performing batch analysis of several data sets, or in any Big Data scenario.

Note that the selection methodology can be performed without the explicit plotting of the LMRD and sample L-skewness/L-kurtosis pairs. However, one can easily ask for the graphical representation of the process, if visualization of the behaviour of the sequential estimation is desired. For illustration purposes, we will graphically expose the process of threshold choice for a simulated sample of $n=500$ points from a hybrid distribution introduced by \cite{northropcoleman2014}.

The density of a Hybrid($u\,,\,\xi$) is continuously defined for positive real values as a constant function up the \emph{true threshold} $u \in (0,1)$ -- i.e., Uniform(0,1) density up to its $u\times100\%$ quantile -- and a GP density with shape $-1<\xi<1$ and scale $\sigma_u=1-u$ for the excesses above $u$; this probability density function can then be written as
\begin{equation}\label{hybrid}
    \text{Hybrid}(x\,|\,u\,,\,\xi)= \begin{cases} 
                1\,,  & 0<x<u\,,\\
                \left(1+\xi\,\frac{x-u}{1-u}\right)^{-\frac{1}{\xi}-1} \,, & u\leq x < x^\uparrow\,,
            \end{cases}
\end{equation}
where $x^\uparrow$ is the distribution's (finite or infinite) right endpoint, that is
\begin{equation*}\label{hybrid_endpoint}
    x^\uparrow:=\sup\{x: \text{Hybrid}(x)<1\} =\begin{cases} 
                \frac{(1+\xi)\,u-1}{\xi}\,,  & \mbox{for} \;\; \xi<0\,,\\
                +\infty\,, & \mbox{for} \;\; \xi>0\,.
            \end{cases}
\end{equation*}
Using such a distribution has one major advantage in this context: the true value of the threshold that the method should return is known, since the hypothesis of a GPd distribution for excesses is true above $u$. Usefulness of this property for assessing the validity of our method will be apparent.

\begin{figure}[ht!]
  \centering
   \includegraphics[width=0.61\textwidth]{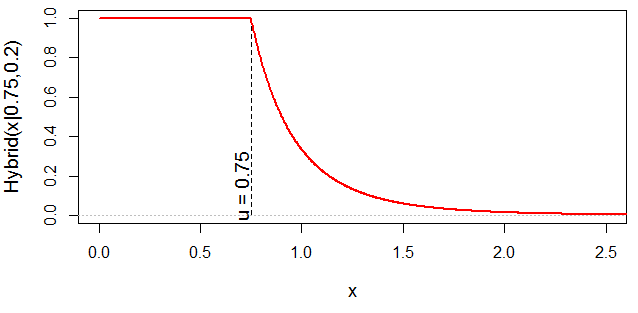}
    \vspace{-0.2cm}
    \caption{\small{Probability density function of the Hybrid($0.75,\,0.2$) distribution.}}\label{fig:hybrid}
\end{figure}

\begin{figure}[ht!]
  \centering
  \vspace{0.7cm}
    \includegraphics[width=0.6\textwidth]{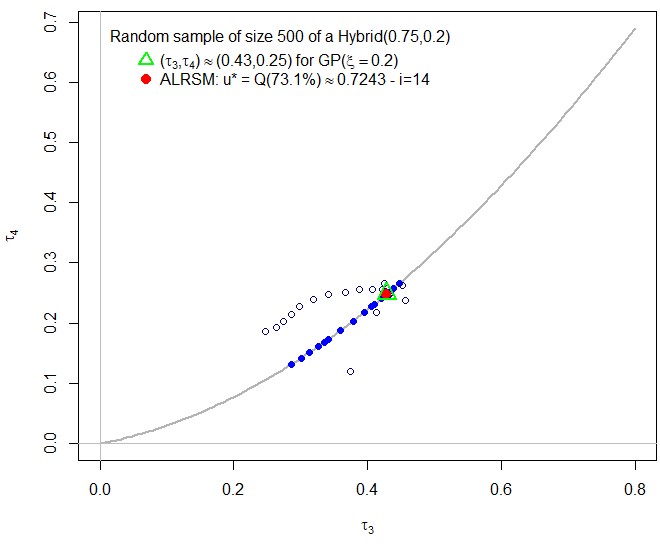}
    \vspace{-0.2cm}
    \caption{\small{Automatic selection procedure for a simulated sample of size $n=500$ from a Hybrid($0.75,\,0.2$) distribution, with sample quantile threshold candidates $\left\{u_i\right\}_{i=1}^{20}$, as in \eqref{I20}. Theoretical L-skewness and L-kurtosis for GPd with same $\xi=0.2$ plotted as hollow triangle; sample estimates $(\mathnormal{t}_{3,u_i},\mathnormal{t}_{4,u_i})\,,\,i=1,\ldots,20$, plotted as hollow circles, with corresponding minimum-distance points on the curve as solid circles; $(\mathnormal{t}_{3,u^*},\mathnormal{t}_{4,u^*})$ for selected threshold as solid red.}}\label{fig:L-mom_select_plot_ex}
\end{figure}

Figure \ref{fig:hybrid} shows the probability density function of the Hybrid distribution with shape parameter $\xi=0.2$ and true threshold at $u=0.75$ -- the distribution underlying the sample of Figure \ref{fig:L-mom_select_plot_ex}. For this illustration, the largest set of 20 candidate levels for the the ALRSM was used. The selected threshold then corresponds to the 14\textsuperscript{th} candidate, the 73.1\% sample quantile with value $u_{14}\approx0.7243$, close to the true level $u=0.75$ after which observations truly come from a GPd(0.2) -- we can see that the associated L-statistics are very close the theoretical $(\tau_3,\tau_4)\approx(0.43,0.25)$ for the proposed GPd with the true shape $\xi=0.2$.

\vspace{0.3cm}
In the next section we will present a comprehensive simulation study aiming at evaluation and validation of the presented methodology against state-of-the-art suggestions from the previously referred literature.

\section{Comparative Simulation Study}

To assess the validity and evaluate the performance of our proposed method, an extensive simulation study was conducted, including comparison of our methodology against two alternative and competitive procedures found in recent literature: the improved diagnostic plots devised by \cite{northropcoleman2014} and the automated threshold selection via ordered GoF tests presented by \cite{baderetal2018}. Although there was the will to compare these methodologies with the Bayesian technique presented by \cite{leeetal2015}, the computational effort required by the latter, as well as the lack of a clear and direct way to automatize it, makes the process impractical for application in a large scale study such as the one described hereafter. We will revisit the work by these authors in Section 4.

All three compared methods stand on the same principle: a given set of candidate thresholds $u_1 \leq \ldots \leq u_I$ from which to choose a single, most appropriate one for a GPd-POT analysis of extremes data. Let us look a bit further into the competing selection procedures, before presenting the simulation results, for a better understanding of the fundamental differences between the methodologies.

\subsection{Remarks on Northrop and Coleman (2014) -- Score Test Selection Method (STSM)}

\begin{figure}[ht!]
  \centering
    \includegraphics[width=0.6\textwidth]{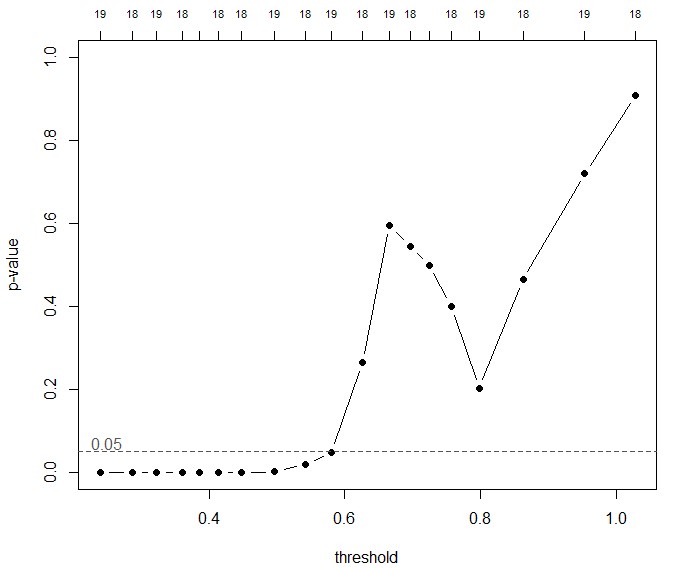}
    \vspace{-0.2cm}
    \caption{\small{Output of R function \emph{score.fitrange}: multiple threshold diagnostic plot for $n=500$ simulated Hybrid($0.75,\,0.2$) observations, with sample quantile threshold candidates $\left\{u_i\right\}_{i=1}^{20}$; dotted horizontal line at significance level $\alpha=0.05$.}}
    \label{fig:northrop_ex}
\end{figure}

Following the work of \cite{wadsworthtawn2012} and in an effort to give higher interpretability to traditional parameter stability plots, \cite{northropcoleman2014} devised a visual diagnostic plot from a multiple-threshold GP model. By considering a piece-wise constant representation of the shape parameter, dependent on the set of $I$ candidate thresholds, 
\begin{equation*}
    \xi(x)= \begin{cases} 
                \xi_i\,,  & u_i<x<u_{i+1}\,,\;\;\text{for } i=1,\ldots,I-1,  \\
                \xi_I \,, & x>u_I\,,
            \end{cases}
\end{equation*}
these authors arrived at a variant of the Generalized Pareto model for which they were able to construct a score test statistic for shape parameter stability. Thus, given the set of null hypotheses $H_0^i: \xi_i=\ldots=\xi_I$, for $i=1,\ldots,I-1$, a plot of the associated p-values can be drawn to show the level after which the constancy of the EVI can be assumed. For illustrative purposes, Figure \ref{fig:northrop_ex} shows the output of the R function \emph{score.fitrange}, kindly made available by the authors, which implements this methodology when applied to the same sample of $n=500$ simulated observations from the Hybrid($0.75,\,0.2$) distribution previously used for illustration of the ALRSM in Figure \ref{fig:L-mom_select_plot_ex}; the same set of $I=20$ sample quantile candidate thresholds in \eqref{I20} was considered.

The problem lies in the interpretation of this p-values' plot: the authors offer the suggestion of viewing the p-values as measure of disagreement between the data and the null hypotheses, such that one should choose to set the threshold at the point where there is approximate stabilization of p-values, or at a point where a sharp increase is visible. For the example of Figure \ref{fig:northrop_ex}, it is not clear what threshold is suggested -- perhaps a choice around the 12\textsuperscript{th} candidate $u_{12}\approx 0.6662$ (the 65.7\% sample quantile) which corresponds to the first peak in the plot. However, if we use the p-values for usual testing at a pre-specified size of 5\%, e.g., non-rejection of the null hypothesis occurs at lower thresholds, leading to the choice of the 10\textsuperscript{th} (borderline decision) or 11\textsuperscript{th} candidates. These candidates correspond, resp., to the 58.3\% and 62.0\% sample quantile, i.e. threshold values $u_{10}\approx 0.5805$ and $u_{11}\approx 0.6258$, which are someway below the true level $u=0.75$.\vspace{0.2cm}

\normalsize{As such, the usage of this methodology is fundamentally subjective, and choice of an automatic implementation is required for our simulation setting. Following a suggestion from the authors, we apply the same selection criterion used by \cite{wadsworth2016}: \emph{the lowest threshold above which the p-value is larger than 0.05 and remains larger than 0.05 at all higher thresholds.} We will refer to this rule as the automatized STSM (aSTSM). In the example illustrated by Figure \ref{fig:northrop_ex}, aSTSM leads to the choice of $u_{11}\approx 0.6258$.  Another alternative would be to select the first level at which the p-value is larger than 0.05, but this does not protect against rejection of the null hypothesis at a higher threshold.}

\subsection{Remarks on Bader et al. (2018) -- Sequential Goodness-of-fit Selection Method (SGFSM)}

Motivated by the results of \cite{choulakian2001} -- regarding the asymptotic distribution of the Anderson-Darling (AD) GoF test statistic for the GPd hypothesis --, and the stopping rules developed by \cite{gselletal2016}, the work of \cite{baderetal2018} proposes an automated threshold selection method built on a sequence of GoF tests with adjustment for error control in a multiple ordered testing scenario.

\newpage
Considering the sequence of ordered null hypotheses  $H_0^i:$\emph{ the distribution of the} $n_i$ \emph{excesses above} $u_i$ \emph{follows the GPd}, for $i=1,\ldots,I$, a rejection rule based on the transformed sequence of p-values from the AD GoF tests is applied -- the ForwardStop rule, given by

\begin{equation} \label{forwardstop}
    \hat{k}_F=\max\left\{ k\in \{1,\ldots,I\}:-\frac{1}{k}\sum_{i=1}^{k}\log(1-p_i)\leq \alpha\right\}
\end{equation}
where $\left\{p_i\right\}_{i=1}^{I}$ is the sequence of raw p-values of the ordered hypotheses. The pre-specified level $\alpha$ is used to control the False Discovery Rate, under the (here violated) assumption of independence of the multiple tests. 
The cutoff $\hat{k}_F$ returned by \eqref{forwardstop} should be interpreted as rejection of all hypotheses $H_0^1,\ldots,H_0^{\hat{k}_F}$.

To apply this selection procedure exactly as described in the referenced paper, we make use of the \emph{gpdSeqTests} and \emph{pSeqStop} functions, implemented by \cite{eva} in their R package \texttt{eva}. Running these functions for the same sample used for the plots in Figures \ref{fig:L-mom_select_plot_ex} and \ref{fig:northrop_ex}, the sequence of raw and adjusted p-values according to the transformation in ForwardStop is given in Table \ref{tab:bader_ex}.

\small{
\begin{table}[ht!]
\begin{center}\small{
\begin{tabular}{ll}
\begin{tabular}{ccccc}
\multirow{2}{*}{$i$}&\multirow{2}{*}{$u_i$}&\multirow{2}{*}{$n_i$}&\emph{AD raw} &\multirow{2}{*}{\emph{ForwardStop}}\\
&&&\emph{p-values}&\\
\toprule
    1     & 0.2388 & 375   & 0.000000 & 0.000000 \\
    2     & 0.2867 & 356   & 0.000000 & 0.000000 \\
    3     & 0.3225 & 338   & 0.000000 & 0.000000 \\
    4     & 0.3597 & 319   & 0.000016 & 0.000004 \\
    5     & 0.3859 & 301   & 0.000003 & 0.000004 \\
    6     & 0.4133 & 282   & 0.000001 & 0.000004 \\
    7     & 0.4477 & 264   & 0.000000 & 0.000003 \\
    8     & 0.4963 & 246   & 0.000018 & 0.000005 \\
    9     & 0.5420 & 227   & 0.001318 & 0.000151 \\
    10    & 0.5805 & 209   & 0.004759 & 0.000613 \\
\bottomrule
\end{tabular}
&\hspace{-0.45cm}
\begin{tabular}{ccccc}
\multirow{2}{*}{$i$}&\multirow{2}{*}{$u_i$}&\multirow{2}{*}{$n_i$}&\emph{AD raw} &\multirow{2}{*}{\emph{ForwardStop}}\\
&&&\emph{p-values}&\\
\toprule
    \textbf{11}    & \textbf{0.6258} & \textbf{190}   & \textbf{0.107317} & \textbf{0.010877} \\
    12    & 0.6662 & 172   & 0.621147 & 0.090855 \\
    13    & 0.6963 & 153   & 0.261141 & 0.107147 \\
    14    & 0.7243 & 135   & 0.533314 & 0.153929 \\
    15    & 0.7572 & 116   & 0.387592 & 0.176358 \\
    16    & 0.7982 & 98    & 0.115800 & 0.173027 \\
    17    & 0.8629 & 79    & 0.482245 & 0.201570 \\
    18    & 0.9539 & 61    & 0.271928 & 0.208002 \\
    19    & 1.0283 & 42    & 0.212107 & 0.209602 \\
    20    & 1.2263 & 24    & 0.031952 & 0.200746 \\
\bottomrule
\end{tabular}
\end{tabular}
}\end{center}
\vspace{-0.2cm}
\caption{\small{Output of functions \emph{gpdSeqTests} and \emph{pSeqStop}: raw and adjusted p-values for the sequential AD GoF tests to a sample of $n=500$ simulated Hybrid($0.75,\,0.2$) observations, with sample quantile threshold candidates $\left\{u_i\right\}_{i=1}^{20}$, defined by \eqref{I20}; the cutoff $\hat{k}_F=11$,  in bold, leads to $u_{12}\approx0.6662$.}}
\label{tab:bader_ex}
\end{table}
}
 
\normalsize{
By directly applying \eqref{forwardstop} we find that the threshold selected by this method is the 12\textsuperscript{th} candidate, corresponding to the 65.7\% sample quantile, a choice similar to the ones suggested by the subjective use of STSM. Again, this level $u_{12}\approx0.6662$ is below the true $u=0.75$. Note that according to the raw AD p-values, one could advocate for the choice of the even lower candidate $u_{11}\approx0.6258$ at a 5\% significance level; however, ForwardStop dilutes this higher p-value with the ones from the previous candidates, leading to a clear rejection of $H_0^{11}$ at this significance.

A small codding intervention was necessary to truly automatize the selection using the outputs of the mentioned functions, making the process appropriate for our simulation study.} 

\subsection{Simulation Scheme}

\normalsize{The performance of the three competing methods was evaluated mainly on random samples simulated from the previously described Hybrid distribution in \eqref{hybrid}. This allows the comparison of methods in terms of mean bias and root mean squared error (RMSE) of the selected threshold, along with that of other estimated parameters and quantiles, since, as referred, the true value of the threshold that the selection methods should return is known for this distribution. Parallel studies (not presented in this work) were conducted on samples generated from the pure GPd and Fr\'{e}chet distribution, but due to the nonexistence of a true cutoff level for these models, interpretability and assessment of selection results is much more challenging. Therefore, our efforts were focused on the Hybrid distribution described, with the advantages it entails.
}

For each sample, all three selection procedures were applied. After the choice of ``optimal'' level $u^*$, a typical POT analysis was conducted -- a summarized description of the usual procedure can be found in Section 3 of \cite{lomba2016} and references therein. Estimation of parameters of the GPd fit to the samples of excesses was performed by maximum likelihood (ML) across the three methodologies to ensure that the comparisons drawn relate only with the threshold selection itself and not to the parameter estimation method, although, as previously mentioned, the expressions in \eqref{lambda1e2etau3GP} and \eqref{relationtauGP} allow for direct estimation of the GP parameters in the L-moment framework. The ML estimates $\hat{\xi}$ and $\hat{\sigma}_u$ were then used to estimate extreme quantiles (return levels) of exceedance probabilities $p=0.01$ and $p=0.001$ 
given by
\begin{equation}\label{quantilestimator}
    \widehat{\chi_p}=u^*+ \frac{\hat{\sigma}_u}{\hat{\xi}}\left(\left(\frac{n\times p}{n^*}\right)^{-\Hat{\xi}}-1\right)\,,
\end{equation}
where $n$ is the complete sample size and $n^*$ the size of the sample of excesses above the selected $u^*$. 

Mean bias and RMSE for the selected threshold and estimated entities were computed from a set of 1500 samples generated according to each combination of several proposed scenarios:
\begin{itemize}
        \item GP tail shape parameter: $\xi=-0.2$;  $\xi=0.2$; $\xi=0.5$; 
        \item True threshold: $u=0.75$; $u=0.5$;
        \item Sample size: $n=1000$; $n=500$; $n=200$.
\end{itemize}

Moreover and as mentioned in Section 2.2, we considered two sets of candidate thresholds of different sizes (10 or 20 candidates, established in accordance with \eqref{I10} and \eqref{I20}, respectively). Thus, we are able to evaluate the performance of the methodologies across sample size, number of candidates and distributional form. Note that scenarios of both heavy and light tails were used.

\subsection{Main Results}

\begin{figure}[ht!]
  \hspace{-0.2cm}
  \begin{subfigure}[]{0.535\textwidth}
            \includegraphics[width=\textwidth]{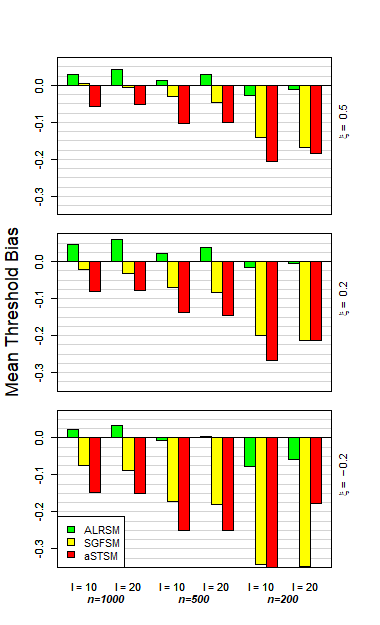}
              \vspace{-1cm}
            \caption{}\label{fig:meanthresholdbias}
  \end{subfigure}
  \hspace{-0.5cm}
  \begin{subfigure}[]{0.535\textwidth}
            \includegraphics[width=\textwidth]{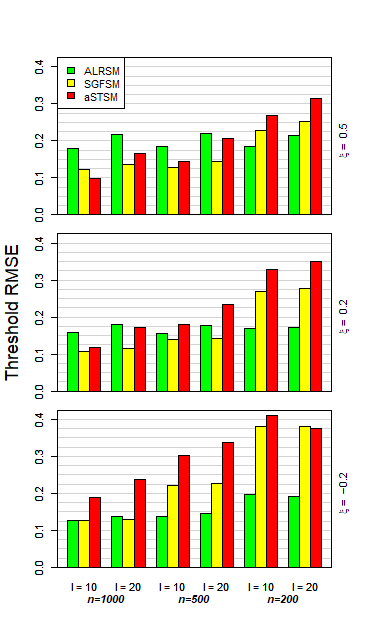}
              \vspace{-1cm}
            \caption{}\label{fig:rmsethreshold}
  \end{subfigure}
  \vspace{-0.2cm}
    \caption{\small{(a) Mean threshold selection bias and (b) RMSE of threshold selection for the 1500 simulations: $n=1000,\,500,\,200$ observations from a Hybrid($0.75,\,\xi$), with $\xi=-0.2,\,0.2,\,0.5$ and sample quantile threshold candidates $\left\{u_i\right\}_{i=1}^{I}$, $I=10,\,20$.}}
\end{figure}

\normalsize{In this section we limit our exposition to the results from scenarios pertaining to the true threshold $u=0.75$, for convenience of presentation. We found that when the median of the distribution is used as the true threshold ($u=0.5$), the comparative performance of the Automatic L-moment Ratio Selection Method against both aSTSM and SGFSM is similar or more favorable than the case  $u=0.75$ presented hereafter.

We summarize the results from this simulation study in the bar plots of Figures \ref{fig:meanthresholdbias} though \ref{fig:rmseq999}, where the colour scheme is green for the ALRSM, red for the aSTSM and yellow for the SGFSM. We assess and compare the performance of these methods in terms of mean bias and RMSE of threshold selection and of EVI estimation, as well as mean ratio of estimates of the 99\% and 99.9\% distributional quantiles' by their true values. Moreover, we look at time efficiency and number of failures in threshold selection or model fitting in each set of 1500 samples.}
 
 \vspace{0.3cm}
\noindent\underline{Threshold Selection}: Looking at Figure \ref{fig:meanthresholdbias} we see that for all considered values of $\xi$, the ALRSM outperforms both alternatives regarding mean bias in most scenarios, with SGFSM being the direct competition, given that the aSTSM presents the largest bias across practically every case considered; for heavy tails and the largest sample size used, the SGFSM presents smaller bias than the L-moment based methodology, but this changes when sample size is reduced or a light tail is considered. A global evaluation of Figure \ref{fig:meanthresholdbias} would suggest that the ALRSM is generally preferable, with a largest mean bias of around 0.075 (in absolute value), against worst case scenarios of mean bias larger than 0.3 in absolute value for both alternatives, almost half the true value of the threshold; additionally, we see that reduction in sample size has less effect on our method -- for heavy tails, smaller sample sizes even showed reduction in selection bias --, while significantly increasing the mean bias of threshold for SGFSM and aSTSM. Finally, the number of candidates considered does not seem to greatly influence any of the performances.

Figure \ref{fig:rmsethreshold} shows the RMSE of the threshold choice. Regarding this indicator, the methods show more comparable performance throughout: the ALRSM yields lower values of RMSE for $n=200$ across all three values of the EVI; for the light tailed distribution, our method is only slightly outperformed by the SGFSM for the largest sample size and candidates' set, while showing, in the smallest sample size scenarios, RMSE values approx. half than those for the alternatives; for the heavy tailed distributions it is outperformed by the SGFSM, as well as by the aSTSM, when the largest sample sizes are used.

\begin{figure}[ht!]\vspace{-0.6cm}
  \hspace{-0.2cm}
  \begin{subfigure}[]{0.535\textwidth}
            \includegraphics[width=\textwidth]{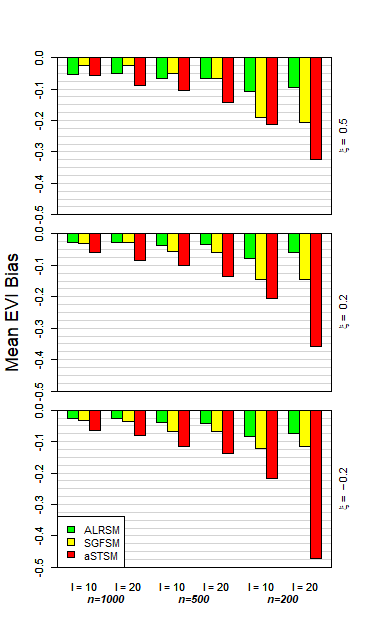}
              \vspace{-1cm}
            \caption{}\label{fig:meanevibias}
  \end{subfigure}
  \hspace{-0.5cm}
  \begin{subfigure}[]{0.535\textwidth}
            \includegraphics[width=\textwidth]{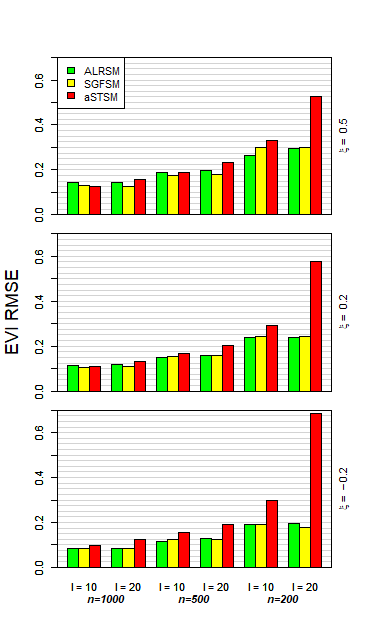}
              \vspace{-1cm}
            \caption{}\label{fig:rmseevi}
  \end{subfigure}
  \vspace{-0.2cm}
    \caption{\small{(a) Mean bias and (b) RMSE of EVI estimation for the 1500 simulations: $n=1000,\,500,\,200$ observations from a Hybrid($0.75,\,\xi$), with $\xi=-0.2,\,0.2,\,0.5$ and sample quantile threshold candidates $\left\{u_i\right\}_{i=1}^{I}$, $I=10,\,20$.}}
\end{figure}

\vspace{0.3cm}
\normalsize{\noindent\underline{EVI estimation}: Recall that estimation of the EVI was performed by ML according to the GPd fit above the level selected for each sample by the methodologies at hand. Mean estimation bias for $\xi$ is always negative, as shown in Figure \ref{fig:meanevibias} -- there is underestimation of the tail heaviness by all methods. The most striking information we can draw from this plot is that estimation of the EVI from the aSTSM is always significantly more biased than the alternatives. Furthermore, the estimation performed according to the ALRSM's threshold is generally the less biased out of the three methods, especially for the lighter tails and smaller sample sizes; it is once again only outperformed by the SGFSM selection in three heavy tail/large sample instances. For a practical situation where the EVI is unknown, this suggests the ALRSM would be preferable.}

\newpage

\normalsize{In Figure \ref{fig:rmseevi} we can see the values of RMSE from the EVI estimation, which appear to be quite uniform across the three methods and almost all simulation scenarios; mind that smaller samples are related to larger values of RMSE for all methods, as is to be expected, but the aSTSM has here exceptionally bad behaviour, with very high RMSE for $I=20$ candidates tested on samples of $n=200$ observations. Otherwise, between the ALRSM and SGFSM, neither method presents clearly lower RMSE for EVI estimation than the other.}

\newpage

\begin{figure}[ht!]\vspace{-0.6cm}
  \hspace{-0.2cm}
  \begin{subfigure}[]{0.535\textwidth}
            \includegraphics[width=\textwidth]{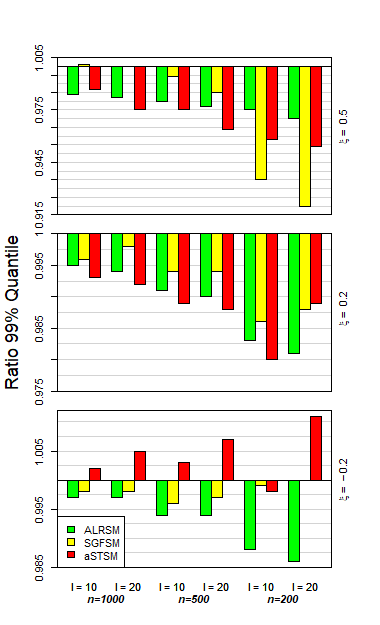}
              \vspace{-1cm}
            \caption{}\label{fig:ratio99}
  \end{subfigure}
  \hspace{-0.5cm}
  \begin{subfigure}[]{0.535\textwidth}
            \includegraphics[width=\textwidth]{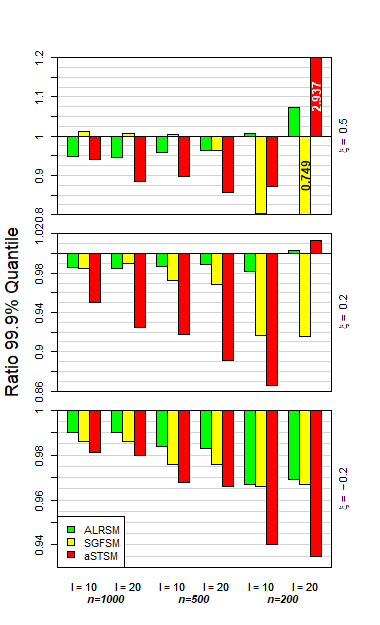}
              \vspace{-1cm}
            \caption{}\label{fig:ratio999}
  \end{subfigure}
  \vspace{-0.2cm}
    \caption{\small{ Mean ratio of estimates over (a) true 99\% quantile $\left(\sfrac{\widehat{\chi_{0.01}}}{\chi_{0.01}}\right)$ and (b) true 99.9\% quantile $\left(\sfrac{\widehat{\chi_{0.001}}}{\chi_{0.001}}\right)$ for the 1500 simulations: $n=1000,\,500,\,200$ observations from a Hybrid($0.75,\,\xi$), with $\xi=-0.2,\,0.2,\,0.5$ and sample quantile threshold candidates $\left\{u_i\right\}_{i=1}^{I}$, $I=10,\,20$.}}
\end{figure}

\begin{figure}[ht!]\vspace{-0.5cm}
  \hspace{-0.2cm}
  \begin{subfigure}[]{0.535\textwidth}
            \includegraphics[width=\textwidth]{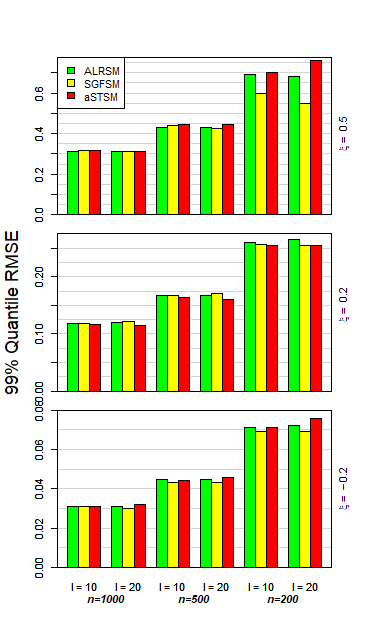}
              \vspace{-1cm}
            \caption{}\label{fig:rmseq99}
  \end{subfigure}
  \hspace{-0.5cm}
  \begin{subfigure}[]{0.535\textwidth}
            \includegraphics[width=\textwidth]{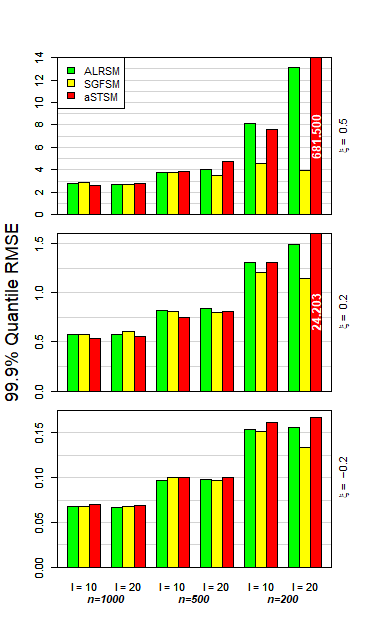}
              \vspace{-1cm}
            \caption{}\label{fig:rmseq999}
  \end{subfigure}
  \vspace{-0.2cm}
    \caption{\small{RMSE of (a) 99\% quantile and (b) 99.9\% quantile estimation for the 1500 simulations: $n=1000,\,500,\,200$ observations from a Hybrid($0.75,\,\xi$), with $\xi=-0.2,\,0.2,\,0.5$ and sample quantile threshold candidates $\left\{u_i\right\}_{i=1}^{I}$, $I=10,\,20$.}}
\end{figure}

\vspace{0.1cm}
\normalsize{\noindent\underline{Extreme quantile estimation}: According to the assumed GPd, quantiles of exceedance probability $p$ are given by \eqref{quantilestimator} -- results presented hereafter were computed from this expression with the threshold selected by each method and associated ML estimates of the distributional shape and scale parameters. Let us look at the mean ratios of estimates over true quantile values, as well as RMSE of estimation of the GPd tail quantiles $\chi_{0.01}$ and $\chi_{0.001}$. We choose here to present the ratio indicator in detriment of the estimation mean bias for better comparability of results from different tail weights, since the true values of the quantiles differ greatly with changes in the underlying EVI.

Figures \ref{fig:ratio99} and \ref{fig:ratio999} show, respectively, the mean ratio values of the estimates over true quantile for $\chi_{0.01}$ and $\chi_{0.001}$. A general appreciation shows a tendency for underestimation, given that mean ratio values fall mostly below the target value 1. Visually obvious exceptions to this remark come from the estimation associated with the aSTSM, which once again is mostly outperformed by the alternatives, in the sense that estimates fall further away from the true values, in average.}

\newpage

\normalsize{Regarding estimation of $\chi_{0.01}$, Figure \ref{fig:ratio99} shows that the SGFSM seems to provide more accurate estimates than the proposed ALRSM for most cases. However, looking at Figure \ref{fig:ratio999}, our conclusion changes with the consideration of the more extreme quantile $\chi_{0.001}$ -- estimation from the ALRSM choice is closer to the true value of the 99.9\% quantile for most scenarios than both alternatives.

Performance of our proposed selection methodology is generally best for smaller values of the EVI. Regardless, the differences in the results of estimation associated with the other methods are clearer for the smallest sample size here considered. Particularly, in Figure \ref{fig:ratio999} we can see that when estimating the quantile of exceedance probability 0.001 with underlying $\xi=0.5$ from samples of size $n=200$ and $I=20$ candidate thresholds, both alternative methodologies to the ALRSM yield significantly inaccurate estimates (the mean ratios fall outside the bounds of the plot). Moreover, the aSTSM and SGFSM results seem to be more sensitive to the decrease in exceedance probability of the quantile, as well as in the tail index value.}

\normalsize{Figures \ref{fig:rmseq99} and \ref{fig:rmseq999} plot the RMSE of both quantiles' estimation, and we can see that all three processes show very similar and comparable performances to each other, across the majority of scenarios. Values of RMSE are higher, as is to be expected, for heavier tails, larger quantiles and smaller sample sizes. This indicator does little to point towards a preferable methodology.}

\vspace{0.3cm}
\noindent\underline{Computational efficiency}: Given that one of our main goals is to be able to objectively perform fast and accurate estimation in a context of batch analysis, it is pertinent to analyze the run-time and process failure count of the competing methods here presented. 

Regarding time efficiency, the L-moment based method we suggest is clearly preferable to the alternatives -- looking at Table \ref{tab:runtime} we see the time, in seconds, of threshold selection and parameter/quantile estimation for the 1500 simulated samples in each scenario, and the difference between the ALRSM and the alternatives is striking. The method took at most 13 seconds to perform the complete analysis of 1500 simulations. Although we have concluded that the SGFSM is the most comparable method to the ALRSM in terms of threshold selection and parameter estimation bias and RMSE, here we are faced with one of its biggest disadvantages: it is the least time efficient method of the three, taking as much as 479 seconds to run 1500 simulations (for the same scenario of $n=200$ observations samples from the Hybrid($0.75,\,-0.2$) and 10 candidate thresholds tested, the ALRSM and aSTSM take 12 seconds and 275 seconds to run 1500 simulations).

\small{
\begin{table}[ht!]
  \centering \small{
    \begin{tabular}{ccc c||c c||c c}
          & & \multicolumn{2}{ c || }{$\mathbf{n=1000}$} & \multicolumn{2}{c||}{$\mathbf{n=500}$} & \multicolumn{2}{c}{$\mathbf{n=200}$} \\
          &       & $I=10$  & $I=20$  & $I=10$  & $I=20$  & $I=10$  & $I=20$ \\
    \toprule
\multirow{3}{*}{$\xi=0.5$} & ALRSM & \textbf{8}     & \textbf{12}   & \textbf{7}     & \textbf{11}    & \textbf{6}     & \textbf{10} \\
\cline{2-8}          & SGFSM & 213   & 442   & 215   & 416   & 224   & 454 \\
\cline{2-8}          & aSTSM & 84    & 334   & 69    & 295   & 57    & 262 \\
    \midrule
    \midrule
\multirow{3}{*}{$\xi=0.2$} & ALRSM & \textbf{8}     & \textbf{13}   & \textbf{7}    &\textbf{11}    & \textbf{6}    & \textbf{10} \\
\cline{2-8}          & SGFSM & 206   & 341   & 199   & 354   & 232   & 433 \\
\cline{2-8}          & aSTSM & 99    & 339   & 76    & 304   & 61    & 271 \\
    \midrule
    \midrule
\multirow{3}{*}{$\xi=-0.2$} & ALRSM & \textbf{8}     & \textbf{12}    & \textbf{6}     & \textbf{11}    & \textbf{7}     & \textbf{12} \\
\cline{2-8}          & SGFSM & 239   & 478   & 233   & 456   & 245   & 479 \\
\cline{2-8}          & aSTSM & 110   & 362   & 77    & 305   & 63    & 275 \\
    \bottomrule
    \end{tabular}%
  }  \caption{\small{Total run time, in seconds, of threshold selection and parameter estimation for 1500 simulated samples with size $n=1000,\,500,\,200$ from a Hybrid($0.75,\,\xi$), with $\xi=-0.2,\,0.2,\,0.5$ and sample quantile threshold candidates $\left\{u_i\right\}_{i=1}^{I}$, $I=10,\,20$.}}\label{tab:runtime}%
\end{table}%
}

\normalsize{Table \ref{tab:failure} shows the number of failed analyses out of the 1500 simulations in each described scenario. It is clear that the SGFSM presents a much higher number of failures than the alternatives, which increases with lightness of the tail (decreasing EVI), number of threshold candidates considered and reduction of sample size. These failures for the SGFSM correspond to fundamental failure of threshold selection, which means that function \emph{gpdSeqTests} failed to produce the sequence of raw p-values from the AD GoF tests. This issue could perhaps be fought with the usage of the bootstrap option included in the function, but it would certainly result in even higher computational times that already are not competitive enough face the ALRSM. On the other hand, the failure count for the ALRSM corresponds to failure of the GPd fit \emph{after} the choice of threshold -- we are able to obtain information regarding the most appropriate threshold so the failure could be fought by considering a different fitting process. It is also for this reason that the aSTSM presents absolutely no failures in the complete study, given that the ML estimation of parameters is performed intrinsically by the \emph{score.fitrange} and not externally by a GPd fitting function.}

\small{
\begin{table}[ht!]
  \centering \small{
    \begin{tabular}{ccc c||c c||c c}
          &       & \multicolumn{2}{c||}{$\mathbf{n=1000}$} & \multicolumn{2}{c||}{$\mathbf{n=500}$} & \multicolumn{2}{c}{$\mathbf{n=200}$} \\
          &       & $I=10$  & $I=20$  & $I=10$  & $I=20$  & $I=10$  & $I=20$ \\
    \toprule
\multirow{3}{*}{$\xi=0.5$} & ALRSM & 0     & 0     & 0     & 0     & 1     & 0 \\
\cline{2-8}          & SGFSM & 55    & 98    & 85    & 197   & 341   & 678 \\
\cline{2-8}          & aSTSM & 0     & 0     & 0     & 0     & 0     & 0 \\
    \midrule
    \midrule
\multirow{3}{*}{$\xi=0.2$} & ALRSM & 0     & 0     & 0     & 0     & 6     & 2 \\
\cline{2-8}          & SGFSM & 232   & 442   & 186   & 326   & 216   & 511 \\
\cline{2-8}          & aSTSM & 0     & 0     & 0     & 0     & 0     & 0 \\
    \midrule
    \midrule
\multirow{3}{*}{$\xi=-0.2$} & ALRSM & 13    & 6     & 16    & 12    & 53    & 37 \\
\cline{2-8}          & SGFSM & 70    & 100   & 111   & 206   & 584   & 869 \\
\cline{2-8}          & aSTSM & 0     & 0     & 0     & 0     & 0     & 0 \\
    \bottomrule
    \end{tabular}%
  } \caption{\small{Total failure count for 1500 simulated samples with size $n=1000,\,500,\,200$ from a Hybrid($0.75,\,\xi$), with $\xi=-0.2,\,0.2,\,0.5$ and sample quantile threshold candidates $\left\{u_i\right\}_{i=1}^{I}$, $I=10,\,20$.}}\label{tab:failure}%
\end{table}%
}

\vspace{0.3cm}
\normalsize{
In summary, we can draw from the presented study the following general observations:
\begin{itemize}
    \item The aSTSM presents itself as the least agreeable method, and would not be recommended against the considered alternatives; 
    \item The suggested ALRSM clearly stands out against the alternatives regarding computational efficiency, being the most appropriate method for batch analysis of large collections of data samples -- complete analysis of 1500 simulated samples takes around 10 seconds; regardless of computational efficiency and accuracy for large amounts of data, good estimation through the ALRSM does not require large individual sample sizes;
    \item The proposed methodology presents the best overall results in terms of pure threshold selection and estimation of the EVI and 99.9\% GPd quantile; it performs satisfyingly well across most considered scenarios, with emphasis on results from smaller samples;
    \item The most competitive alternative to the ALRSM, in terms of estimation accuracy, is the SGFSM, occasionally outperforming the former; however, in most such cases there is only a slight difference in performances, and this should be weighted by the heavier computational intensity and higher failure rate of the SGFSM.
\end{itemize}}

\section{Examples: Significant Wave Height Data Sets}

We will now illustrate the behaviour of our proposed threshold selection technique against other existing classical tools and   benchmark  methodologies in two real world data sets previously addressed in \cite{northropetal2017} -- our baseline reference hereafter. The series of hindcasts of storm peak significant wave heights from two unnamed locations in the Gulf of Mexico (GoM) and in the North Sea (NS) are publicly available in the R package \texttt{threshr} by \cite{threshr}, and have been treated so that observations can be fairly considered independent and identically distributed . 

Both data series have different expected behaviours, since the natural conditions that result in such registered storm peaks are inherently different. Every winter several tens of storms impact sea conditions in the North Sea, whereas in the Gulf of Mexico severe sea states are normally caused by hurricanes formed in the Atlantic Ocean, which occur more rarely. Thus, regardless of the larger number of events registered for the NS location, potentially higher significant wave heights (s.w.h.) are expected to be seen during storms in the GoM. As such, we expect that an extreme value analysis can detect these underlying differences and provide appropriate inference on parameters and return levels for each environment.

\begin{figure}[h!]
  \hspace{-0.45cm}
  \begin{subfigure}[]{0.535\textwidth}
            \includegraphics[width=\textwidth]{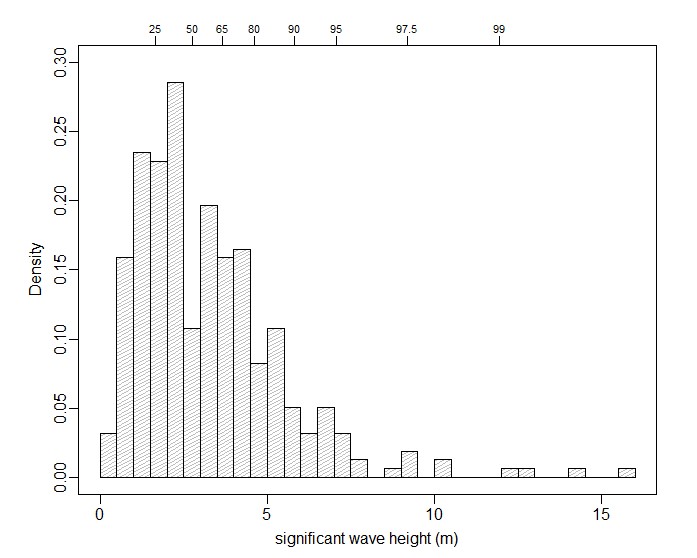}
  \end{subfigure}
  \hspace{-0.55cm}
  \begin{subfigure}[]{0.535\textwidth}
            \includegraphics[width=\textwidth]{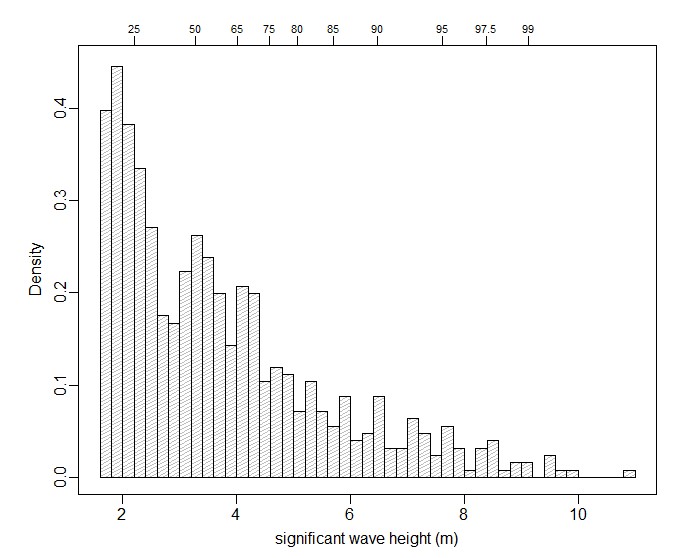}
  \end{subfigure}
  \vspace{-0.2cm}
    \caption{\small{Histograms for the hindcasts of storm peaks significant wave heights data sets, with sample quantiles indicated on the upper axis scale; \emph{Left:} 315 obs. from the GoM; \emph{Right:} 628 obs. from the NS.}}\label{fig:histograms}
\end{figure}

Figure \ref{fig:histograms} shows the histograms for both significant wave height sets (in meters): on the left plot, 315 storm peak s.w.h. registered in one location in the GoM from September 1900 to September 2005, with an average of 3 yearly observations; on the right, 628 storm peak s.w.h. registered in the winter periods (October through March) from October 1964 to March 1995 in a norther NS location, averaging $\approx 20.26$ entries per year.

The following sections summarize the GPd-POT analysis performed on these samples, aiming at comparison of threshold selection by several methods, and drawing inference regarding 100, 1000 and 10 000 year return levels (in meters).

\subsection{Gulf of Mexico}

In their analysis of the 315 observations of the GoM data set, \cite{northropetal2017} present some considerations regarding threshold choice:
\begin{itemize}
    \item From the sample's histogram, where the apparent mode of the data is shown to be above the first sample quartile, thresholds below this level are judged to be inappropriate, given that the GPd has its mode at the origin;
    \item From the stability plot of the ML estimates of $\xi$ versus threshold, seen in Figure 2.(c) of the reference, approximate stability is judged to occur around the 70\% sample quantile;
    \item From the p-value plot given by the STSM (using 20 candidate thresholds from 0 to 90\% sample quantiles in steps of 5\%) seen in Figure 2.(d) of the reference, a sharp increase in the p-values indicates a choice of threshold around the 55\% sample quantile;
    \item From their proposed Bayesian single threshold selection methodology, thresholds in the region of the 60-70\% sample quantiles are suggested, while candidates above the 95\% sample quantile are confirmed to be undesirable (very few excesses can be counted above this level). 
\end{itemize}

\normalsize{Before using our own ALRSM to choose an appropriate threshold for this data set, we investigated how 5 other known methodologies fared in this task, registering the corresponding selected thresholds and estimating the underlying tail index and 100, 1000 and 10000 year return levels -- given the average of 3 observations per year in this data set, these return levels correspond to extremal quantiles of exceedance probabilities $\sfrac{1}{300}$, $\sfrac{1}{3000}$ and $\sfrac{1}{30000}$ respectively. The analysis results are summarized in Table \ref{tab:gom}.}

\small{
\begin{table}[htbp!]
  \centering\small{

    \begin{tabular}{cccccccccc}
    &&&\emph{Sample}&&&&&\\
    \multicolumn{2}{c}{\emph{Method}} & \multicolumn{1}{c}{$I$} & \multicolumn{1}{c}{\emph{quantile (\%)}} & \multicolumn{1}{c}{$u^*$} & \multicolumn{1}{c}{$n^*$} & \multicolumn{1}{c}{$\hat{\xi}_{ML}$} & \multicolumn{1}{c}{$\widehat{RL_{100}}$} & \multicolumn{1}{c}{$\widehat{RL_{1000}}$} & \multicolumn{1}{c}{$\widehat{RL_{10000}}$} \\
    \toprule
    \multicolumn{2}{c}{\textbf{MRLP WLS}} & \textbf{295} & $\approx35.5$ & 2.135 & 203   & -0.027 & 13.24 & 17.63 & 25.64 \\
    \midrule
    \midrule
    \multicolumn{2}{c}{\textbf{GPd QQ-plot}} & \textbf{305} & $\approx 72$ & 4.136 & 89    & 0.195 & 14.85 & 25.35 & 41.80 \\
    \midrule
    \midrule
    \multicolumn{2}{c}{\multirow{2}{*}{\textbf{Surprise Plot}}} & \textbf{10} & 70    & 3.976 & 95    & 0.146 & 14.40 & 23.06 & 35.18 \\
\cmidrule{3-10}    \multicolumn{2}{c}{} & \textbf{20} & 73.1  & 4.182 & 85    & 0.173 & 14.65 & 24.26 & 38.58 \\
    \midrule
    \midrule
    \multicolumn{2}{c}{\multirow{2}{*}{\textbf{aSTSM}}} & \textbf{10} & 55    & 3.160 & 23    & 0.075 & 13.85 & 20.40 & 28.20 \\
\cmidrule{3-10}    \multicolumn{2}{c}{} & \textbf{20} & 25    & 1.660  & 236   & -0.062 & 13.05 & 16.85 & 20.15 \\
    \midrule
    \midrule
    \multicolumn{2}{c}{\multirow{2}{*}{\textbf{SGFSM}}} & \textbf{10} & 55    & 3.160 & 142   & 0.075 & 13.85 & 20.40 & 28.20 \\
\cmidrule{3-10}    \multicolumn{2}{c}{} & \textbf{20} & 54.6  & 3.124 & 143   & 0.063 & 13.75 & 19.99 & 27.19 \\
    \midrule
    \midrule
    \multicolumn{2}{c}{\multirow{3}[6]{*}{\textbf{ALRSM}}} & \textbf{10} & 70    & 3.976 & 95    & 0.146 & 14.40 & 23.06 & 35.18 \\
\cmidrule{3-10}    \multicolumn{2}{c}{} & \textbf{20} & 73.1  & 4.182 & 85    & 0.173 & 14.65 & 24.26 & 38.58 \\
\cmidrule{3-10}    \multicolumn{2}{c}{} & \textbf{305} & $\approx73$ & 4.170  & 86    & 0.179 & 14.70 & 24.53 & 39.37 \\
    \bottomrule
    \end{tabular}%
}    \caption{\small{Threshold selection and inference results for the Gulf of Mexico data set under the six considered selection methodologies.}}\label{tab:gom}%
\end{table}%
}

\begin{figure}[h!]
  \hspace{-0.4cm}
  \begin{subfigure}[]{0.535\textwidth}
            \includegraphics[width=\textwidth]{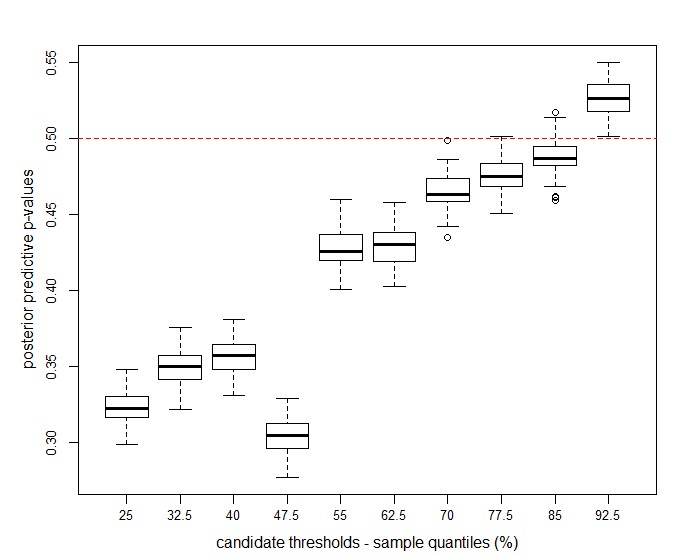}
  \end{subfigure}
  \hspace{-0.5cm}
  \begin{subfigure}[]{0.535\textwidth}
            \includegraphics[width=\textwidth]{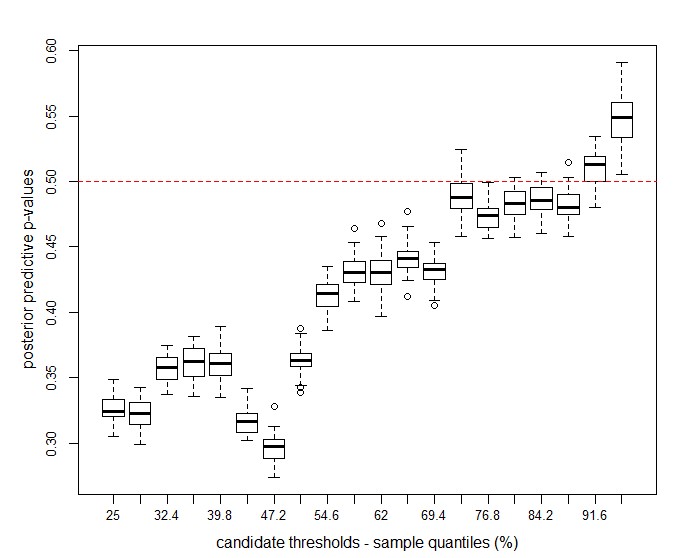}
  \end{subfigure}
  \vspace{-0.2cm}
    \caption{\small{Measures of surprise versus threshold candidates for the Gulf of Mexico data set, for two sets of sample quantile candidate thresholds; \emph{Left:} $I=10$ candidates; \emph{Right:} $I=20$} candidates.}\label{fig:surprise_gom}
\end{figure}

\normalsize{Applying the WLS fitting to the MRLP procedure detailed in \cite{langousis2016} (and previously revised in Section 1), every sample point is considered a possible candidate, apart from the largest 20 observations (as suggested by the authors, we ensure a minimum number of 10 excesses used to compute the mean excess function, and another minimum of 10 points to compute the fit above the last viable candidate). The threshold selected as corresponding to the first locally minimum MSE of the WLS fits is around the 35.5\% sample quantile, with an associated slightly negative estimate of the EVI. On the other hand, if we look for the threshold that, among the first 305 sample points (excluding the largest 10 points), yields the maximum correlation of the linear fit to the GPd QQ-plot, our selection is much higher, around the 72\% sample quantile. Moreover, the associated EVI estimate is also much larger and clearly positive. The return levels estimated from these selections are strikingly different, making them inconclusive.

The previously mentioned Bayesian method of plotting \emph{surprise measures} from \cite{leeetal2015} was also used on this data set, thanks to the kind concession of the computer code from the authors -- the resulting plots can be seen in Figure \ref{fig:surprise_gom}, where posterior predictive p-values (the surprise measure) for the reciprocal likelihood test statistic are represented for each candidate threshold. According to the authors, p-values around 0.5 are to be interpreted as compatibility with the null model. Since Bayesian MCMC methodologies such as this one are very computationally intensive, only the baseline sets of $I=10$ and $I=20$ sample quantile candidates were considered (with run times of $\approx 11$ and $\approx 22$ minutes for this sample, respectively). For these sets of candidates, the suggested levels from a subjective visual analysis of the plots are the 70\% and 73.1\% sample quantiles, producing estimates for return levels that differ at most in 3.4 meters for $RL_{10000}$.}

\begin{figure}[ht!]
  \hspace{-0.4cm}
  \begin{subfigure}[]{0.535\textwidth}
            \includegraphics[width=\textwidth]{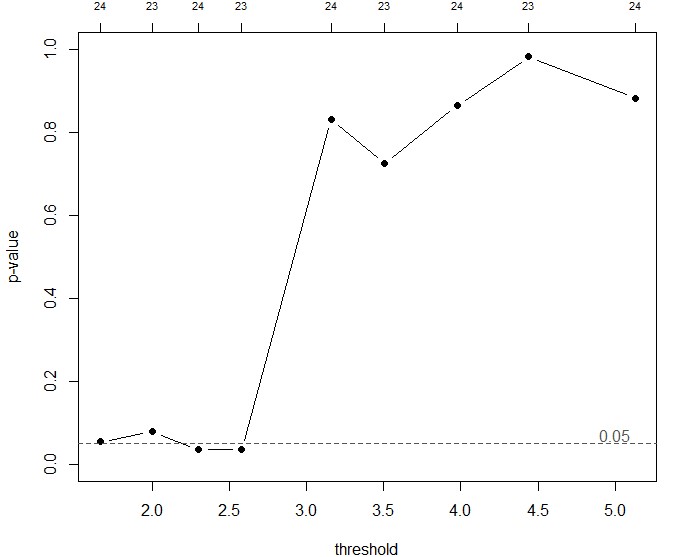}
  \end{subfigure}
  \hspace{-0.5cm}
  \begin{subfigure}[]{0.535\textwidth}
            \includegraphics[width=\textwidth]{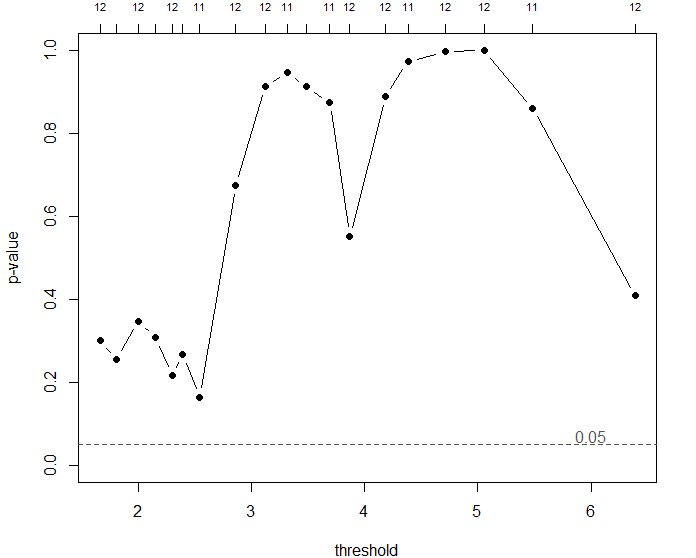}
  \end{subfigure}
  \vspace{-0.2cm}
    \caption{\small{p-value plots versus threshold candidates of the aSTSM for the Gulf of Mexico data set, for two sets of sample quantile candidate thresholds; \emph{Left:} $I=10$ candidates; \emph{Right:} $I=20$} candidates.}\label{fig:northrop_gom}
\end{figure}

\normalsize{The methodologies used for the simulation study, the sTSTM and SGFSM, present close results for this data set, with the exception of the sTSTM applied to testing the 20 sample quantile candidates. Once again, only the two usual sets of $I=10$ and $I=20$ sample quantile candidates were used, given the efficiency concerns. The p-value plots produced for the aSTSM can be seen in Figure \ref{fig:northrop_gom}. When considering 10 candidates, the choice suggested by the aSTSM is the 55\% sample quantile, the same chosen by the SGFSM in the same conditions; when considering 20 candidates, the SGFSM gives the agreeable level of 54.6\% sample quantile, similar to the $I=20$ case, while the aSTSM diverges from this norm, with all p-values above the nominal level $\alpha=0.05$, thus pointing to the choice of the 25\% sample quantile (which we suspect to be too low). Apart from the latter exception, estimation of the return levels does not differ significantly.}

\vspace{0.4cm}

Finally, we look at results from application of our ALRSM: when the candidates considered are the usual $I=10$ and $I=20$ sample quantiles, results coincide perfectly with the selection (and, consequently, parameter estimation) from the surprise plots of \cite{leeetal2015}. However, results from the ALRSM appear almost instantaneously, whereas the competing method takes significant amounts of time to produce the surprise plots, as mentioned. Furthermore, our selection method, for its efficient nature, allows us to perform a naive analysis where all sample points are considered candidates, for direct comparability with selections driven from the parameter stability plots, WLS MRLP and maximum correlation GPd QQ-plots. This is not the case for the other methodologies: only the aSTSM was able to perform faced with the complete sample candidates (repeated sample values had to be removed from the candidate set), and even so the selection was not satisfying, given that the threshold suggested, after a 15 minute run time, was above the 95\% sample quantile.
The ALRSM, given the same 305 candidates as were used for the GPd QQ-plots, provides us with a chosen threshold around the 73\% sample quantile in just a few seconds -- the choice using this general method thus looks to not be very sensitive to the number of candidates considered.

\vspace{0.3cm}
General conclusions regarding this data set are thusly hard to draw, given that selected thresholds ranging from the 55\% to the 73\% sample quantile each have the support of several methodologies; this translates in significant differences in the estimates for the most extreme quantile considered -- around 13 meters between the best and worst case scenarios for $RL_{10000}$. This also translates the associated range of tail weights that are estimated, from $\hat{\xi}=0.063$ to $\hat{\xi}$ close to 0.2, a considerably heavy tail. However, it is the selection from our ALRSM (supported by the surprise plots' selection and close to the maximum correlation QQ-plot choice) which seems to come in closer accordance to the conclusions from the Bayesian method in \cite{northropetal2017}.

\subsection{North Sea}

The complete approach to this data set is in all steps analogous to the one presented above for the Gulf of Mexico data set. As such, a more abbreviated description style will be employed hereafter. 

In their analysis of the 628 observations of the NS data set, \cite{northropetal2017} present some considerations regarding threshold choice:
\begin{itemize}
    \item Analogously to the previously presented data set, thresholds below the first sample quartile are judged to be unrealistic from analysis of the sample's histogram;
    \item The stability plot of the ML estimates of $\xi$ versus threshold, seen in Figure 2.(a) of the reference, once again points to approximate stability at thresholds around the 70\% sample quantile;
    \item From the p-value plot given by the STSM (using 20 candidate thresholds from 0 to 90\% sample quantiles in steps of 5\%) seen in Figure 2.(b) of the reference, a sharp increase in the p-values indicates a choice of threshold in the region of the 70\% sample quantile;
    \item From their proposed Bayesian single threshold selection methodology, thresholds in the region of the 25-35\% sample quantiles are suggested. 
\end{itemize}

For the data at hand, the 100, 1000 and 10000 year return levels correspond to extremal quantiles of exceedance probabilities $\sfrac{1}{2026}$, $\sfrac{1}{20260}$ and $\sfrac{1}{202600}$ resp., given the average of $\approx20.26$ observations per year. The complete analysis' results for the 6 considered methodologies are summarized in Table \ref{tab:ns}.
\small{
\begin{table}[ht!]
  \centering\small{

    \begin{tabular}{cccccccccc}
    &&&\emph{Sample}&&&&&\\
    \multicolumn{2}{c}{\emph{Method}} & \multicolumn{1}{c}{$I$} & \multicolumn{1}{c}{\emph{quantile (\%)}} & \multicolumn{1}{c}{$u^*$} & \multicolumn{1}{c}{$n^*$} & \multicolumn{1}{c}{$\hat{\xi}_{ML}$} & \multicolumn{1}{c}{$\widehat{RL_{100}}$} & \multicolumn{1}{c}{$\widehat{RL_{1000}}$} & \multicolumn{1}{c}{$\widehat{RL_{10000}}$} \\
    \toprule
    \multicolumn{2}{c}{\textbf{MRLP WLS}} & \textbf{608} & $\approx13$ & 1.907 & 457   & -0.219 & 10.61 & 11.80 & 12.52 \\
    \midrule
    \midrule
    \multicolumn{2}{c}{\textbf{GPd QQ-plot}} & \textbf{618} & $\approx71$ & 4.300 & 180    & -0.328 & 10.76 & 11.25 & 11.47 \\
    \midrule
    \midrule
    \multicolumn{2}{c}{\multirow{2}{*}{\textbf{Surprise Plot}}} & \textbf{10} & 77.5    & 4.809 & 142    & -0.346 & 10.72 & 11.17 & 11.37 \\
\cmidrule{3-10}    \multicolumn{2}{c}{} & \textbf{20} & 73.1  & 4.421 & 169    & -0.338 & 10.73 & 11.20 & 11.41 \\
    \midrule
    \midrule
    \multicolumn{2}{c}{\multirow{2}{*}{\textbf{aSTSM}}} & \textbf{10} & 25    & 2.204 & 470    & -0.256 & 11.02 & 11.73 & 12.13 \\
\cmidrule{3-10}    \multicolumn{2}{c}{} & \textbf{20} & 58.3    & 3.623  & 262   & -0.256 & 11.04 & 11.75 & 12.15 \\
    \midrule
    \midrule
    \multicolumn{2}{c}{\multirow{2}{*}{\textbf{SGFSM}}} & \textbf{10} & 25    & 2.204 & 470    & -0.256 & 11.02 & 11.73 & 12.13 \\
\cmidrule{3-10}    \multicolumn{2}{c}{} & \textbf{20} & 25    & 2.204 & 470    & -0.256 & 11.02 & 11.73 & 12.13 \\
    \midrule
    \midrule
    \multicolumn{2}{c}{\multirow{3}[6]{*}{\textbf{ALRSM}}} & \textbf{10} & 77.5    & 4.809 & 142    & -0.346 & 10.72 & 11.17 & 11.37 \\
\cmidrule{3-10}    \multicolumn{2}{c}{} & \textbf{20} & 80.5  & 5.113 & 123    & -0.355 & 10.71 & 11.14 & 11.33 \\
\cmidrule{3-10}    \multicolumn{2}{c}{} & \textbf{618} & $\approx11$ & 1.870  & 557    & -0.215 & 11.38 & 12.31 & 12.87 \\
    \bottomrule
    \end{tabular}%
}    \caption{\small{Threshold selection and inference results for the North Sea data set under the six considered selection methodologies.}}\label{tab:ns}%
\end{table}%
}

\begin{figure}[ht!]
  \hspace{-0.4cm}
  \begin{subfigure}[]{0.535\textwidth}
            \includegraphics[width=\textwidth]{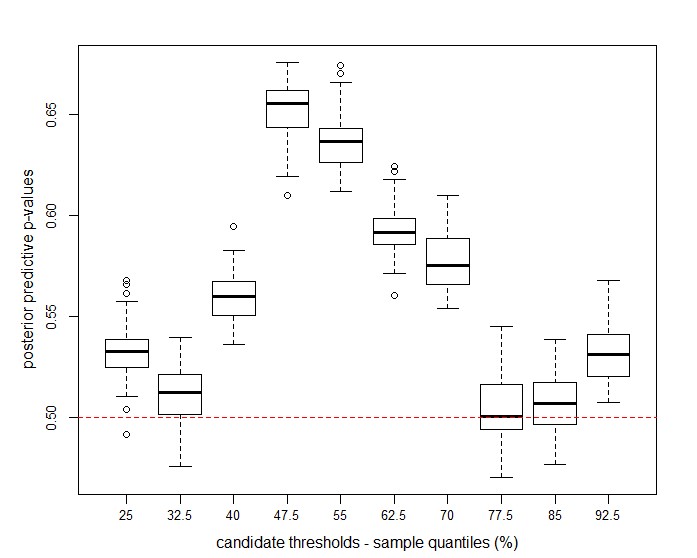}
  \end{subfigure}
  \hspace{-0.5cm}
  \begin{subfigure}[]{0.535\textwidth}
            \includegraphics[width=\textwidth]{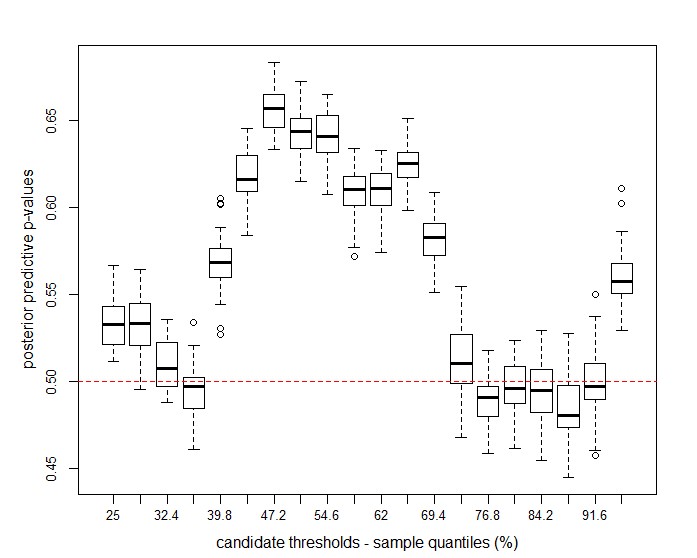}
  \end{subfigure}
  \vspace{-0.2cm}
    \caption{\small{Measures of surprise versus threshold candidates for the North Sea data set, for two sets of sample quantile candidate thresholds; \emph{Left:} $I=10$ candidates; \emph{Right:} $I=20$} candidates.}\label{fig:surprise_ns}
\end{figure}

\normalsize{The minimum MSE of the WLS fits to the MRLP procedure considers here 608 candidates, and the threshold selected is around the 13\% sample quantile, with an associated estimate of the EVI slightly smaller than $-0.2$. In turn, the threshold that, among the first 618 sample points (excluding the largest 10 points), yields the maximum correlation of the linear fit to the GPd QQ-plot, is around the 71\% sample quantile, and the associated EVI estimate is even smaller (a little below $-0.32$). Regardless of these differences, the return levels estimated from these selections are not so divergent, consequence of the lightness of the tail.

The resulting plots of the Bayesian measures of surprise methodology can be seen in Figure \ref{fig:surprise_ns}. For the baseline sets of $I=10$ and $I=20$ sample quantile candidates for this sample, run times of $\approx 17$ and $\approx 32$ minutes were required to draw the surprise plots. For these sets of candidates, the suggested levels are the 77.5\% and 73.1\% sample quantiles, producing estimates of the return levels that differ very slightly -- the largest difference being 0.4 meters for estimates of $RL_{10000}$. Once again, the EVI is estimated to be negative, in both cases below $-0.33$.}

\begin{figure}[H]
  \hspace{-0.4cm}
  \begin{subfigure}[]{0.535\textwidth}
            \includegraphics[width=\textwidth]{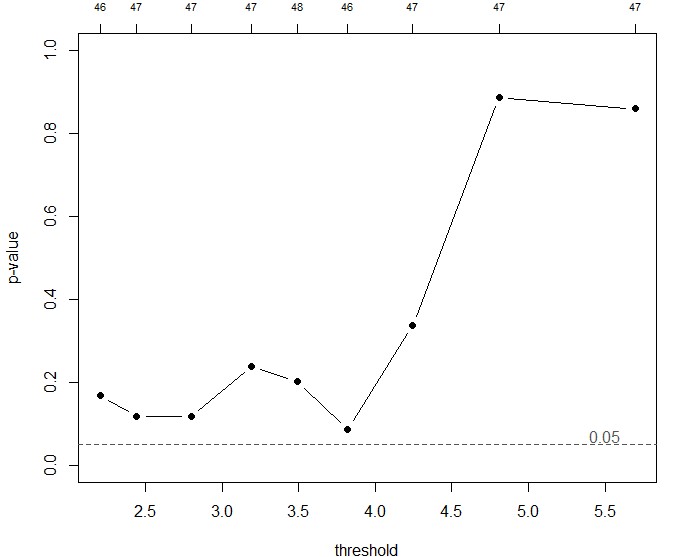}
  \end{subfigure}
  \hspace{-0.5cm}
  \begin{subfigure}[]{0.535\textwidth}
            \includegraphics[width=\textwidth]{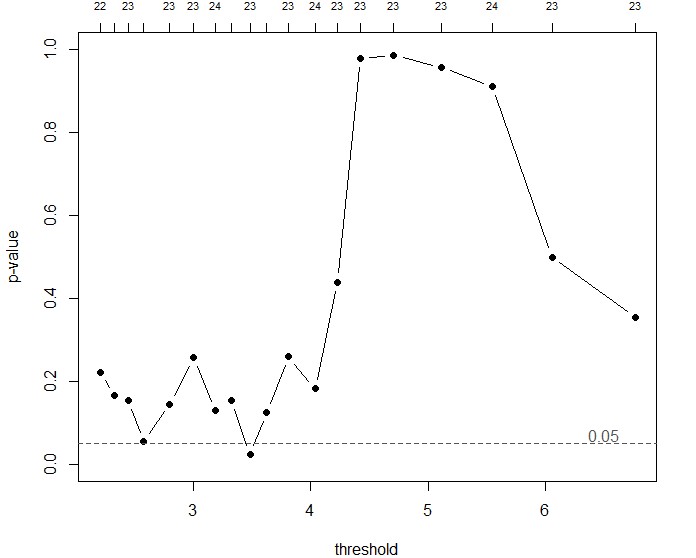}
  \end{subfigure}
  \vspace{-0.2cm}
    \caption{\small{p-value plots versus threshold candidates of the aSTSM for the North Sea data set, for two sets of sample quantile candidate thresholds; \emph{Left:} $I=10$ candidates; \emph{Right:} $I=20$} candidates.}\label{fig:northrop_ns}
\end{figure}

Analogously to the GoM case, the sTSTM and SGFSM present close results for this data set, with the same exception of the sTSTM applied to testing the 20 sample quantile candidates. Once again, only the two usual sets of $I=10$ and $I=20$ sample quantile candidates were used, given the efficiency concerns, even more relevant for this larger data set. The p-value plots produced for the aSTSM can be seen in Figure \ref{fig:northrop_ns}. When considering 10 candidates, the suggested choice by the aSTSM is the lowest candidate, the 25\% sample quantile, and the same choice is indicated by the SGFSM for both sets of candidates; when considering 20 candidates, the aSTSM diverges from this norm, pointing to the choice of the 58.3\% sample quantile. Regardless of the latter exception, estimation of the return levels does not differ up to the first decimal place (estimates are the same to the order of decimeters). 

Finally, we look at results from application of our ALRSM: when the candidates considered are the usual $I=10$ sample quantiles, results coincide perfectly with the selection (and, consequently, parameter estimation) from the surprise plots of \cite{leeetal2015} in the same conditions. Once again, the ALRSM has the advantage of nearly instantaneous results. For the case where $I=20$ candidates are considered, the ALRSM selects a higher threshold at the 80.5\% sample quantile (the highest selection among the presented methods), but this does not greatly change the inference results: return level estimates are the same to the order of decimeters to those associated with the 77.5\% sample quantile threshold. If we once again attempt to naively select a threshold from the complete sample of observations, for direct comparability with the classical visual techniques, the ALRSM provides us with a chosen threshold around the 11\% sample quantile, a much lower level than the ones chosen from the smaller candidate sets -- this yields larger return level estimates, however still not differing more than 1.6 meters from the lowest corresponding estimate. Neither the surprise plot, aSTSM or SGFSM were able to perform in reasonable time facing these conditions.

\vspace{0.3cm}
Alike what happened for the GoM data, there is no indication of a unique consensual threshold choice, as seldom is in practical applications. There is some support from several methods towards choices either in the vicinity of the 25\% or around the 75\% sample quantiles. These two currents are mirrored by the two suggested conclusions of \cite{northropetal2017}, from subjective analysis of the p-value plots and from application of the proposed Bayesian methodology. However, regarding the more general objective of return level estimation, we conclude that differences in level selection do not carry as flagrantly into estimation as for the previous data set. Again, there seems to be some agreement between the ALRSM and the surprise plot methodology, always with the caveat that the latter is much more computationally intensive and requires visual subjective interpretation  -- issues we aimed to eliminate when designing the ALRSM. Moreover, the two other automatic methods, the aSTSM and the SGFSM, also appear to be mostly in agreement.

\section{Concluding Remarks}

Under the GPd-POT framework, the first step to be taken previously to any inference is the choice of an appropriate threshold above which the sample of excesses  can be though of as approximately GP. The influence of this choice carries out to all parts of the POT analysis, so it must be carefully addressed. 

Aiming at objectivity, automation and efficiency, we presented an heuristic spin to a fixed threshold selection procedure built on the well developed theory of L-moments -- the Automatic L-moment Ratio Selection Method. Our purpose is to find an intuitive and reliable methodology that performs well without significant input from its user, useful both in the presence of small samples as well as large collections of data sets, and without need for any visual inspection.

A comprehensive simulation study was conducted, and the particular results presented in this paper were obtained with data generated from a convenient family of distributions -- misspecified below a known true threshold and truly GP distributed above that level. Validity of the methodology was then tested on two real world data sets of significant wave heights. Comparisons were drawn against prominent and competitive methods from the literature, considered, where possible, in their automatized forms.

The ARLSM proposed in this work is shown to perform well both in the simulation context and applied to real data, when compared to the other considered methods from the literature, standing out in terms of threshold selection accuracy and the highly desired computational efficiency, providing results in a few short seconds. The relatively high RMSE associated with some of the parameters estimation is cause for some concern. The method shows robustness against sample size and also against the number of candidate thresholds considered. It also appears to have improved performance for smaller values of the tail index (lighter tails), although still providing acceptable results for the simulations where $\xi=0.5$ -- already considered an unusually heavy tail for most practical situations \citep{clausetetal2009}. The computer code developed for the ARLSM is available from the corresponding author upon reasonable request.

Of course, there is room for improvement of the method and for further work under this setup. Although falling outside the scope of this paper, the following approaches are under study and will be addressed in future work already planned by the authors:
\begin{itemize}
    \item asymptotic properties are known and have been studied in the context of L-moments \citep{hosking1986,hoskingwallis1987}, possibly allowing for a more mathematically rigorous design of the selection procedure;
    \item studying the work of \cite{withersnadarajah2011} regarding bias reduction of estimation of L-skewness and L-kurtosis might lead to improved threshold choices;
    \item a known issue in estimation under the GPd is the effect of data quantization (see e.g. \citealp{deiddapuliga2009}); this topic is extensively addressed by the referred \cite{baderetal2018} and \cite{langousis2016}; submitting the data to a \emph{jittering} process previously to threshold selection should be investigated as a way to improve this methodology.
\end{itemize}
 
It is known that a handicap of choosing a single threshold is the dismissal of a possibly significant uncertainty that should be weighted when it comes to setting a fixed level that will constrain the entire analysis. However, this is still the most straightforward way of instruction practitioners from other scientific fields on how to apply extreme value analysis tools without the requirement of a deeper understanding of the very broad subject of EVT, but still (hopefully) in a sufficiently rigorous and accurate fashion.

\section*{Acknowledgments}
Partial support of the Funda\c{c}\~{a}o para a Ci\^{e}ncia e a Tecnologia, I.P. under doctoral grant SFRH/BD/ 130764/2017 (J.S.L.) and through project UID/MAT/00006/2019 (J.S.L. and M.I.F.A.) is gratefully acknowledged. We would also like to thank Dr. Kate J. E. Lee and Dr. Paul J. Northrop for kindly providing us with the computer code for the referred  alternative methodologies.


\bibliographystyle{apa}
\small\bibliography{references}

\end{document}